\documentclass[ reprint,superscriptaddress, amsmath,amssymb,aps,floatfix]{revtex4-1}
\pdfoutput=1
\usepackage{graphicx}
\usepackage{epstopdf}
\usepackage{color}
\usepackage{dcolumn}
\usepackage{bm}
\usepackage{hyperref}
\usepackage{upgreek} 
\usepackage{siunitx}
\usepackage{booktabs}
\usepackage[normalem]{ulem} 
\usepackage{sverb}
\usepackage{enumitem}

\begin{document}

\title{Signal yields, energy resolution, and recombination fluctuations in liquid xenon}

\author{D.S.~Akerib} 
\affiliation{Case Western Reserve University, Department of Physics, 10900 Euclid Ave, Cleveland, OH 44106, USA}
\affiliation{SLAC National Accelerator Laboratory, 2575 Sand Hill Road, Menlo Park, CA 94205, USA}
\affiliation{Kavli Institute for Particle Astrophysics and Cosmology, Stanford University, 452 Lomita Mall, Stanford, CA 94309, USA}

\author{S.~Alsum} 
\affiliation{University of Wisconsin-Madison, Department of Physics, 1150 University Ave., Madison, WI 53706, USA}

\author{H.M.~Ara\'{u}jo} 
\affiliation{Imperial College London, High Energy Physics, Blackett Laboratory, London SW7 2BZ, United Kingdom}

\author{X.~Bai} 
\affiliation{South Dakota School of Mines and Technology, 501 East St Joseph St., Rapid City, SD 57701, USA}

\author{A.J.~Bailey} 
\affiliation{Imperial College London, High Energy Physics, Blackett Laboratory, London SW7 2BZ, United Kingdom}

\author{J.~Balajthy} 
\affiliation{University of Maryland, Department of Physics, College Park, MD 20742, USA}

\author{P.~Beltrame} 
\affiliation{SUPA, School of Physics and Astronomy, University of Edinburgh, Edinburgh EH9 3FD, United Kingdom}

\author{E.P.~Bernard} 
\affiliation{University of California Berkeley, Department of Physics, Berkeley, CA 94720, USA}
\affiliation{Yale University, Department of Physics, 217 Prospect St., New Haven, CT 06511, USA}

\author{A.~Bernstein} 
\affiliation{Lawrence Livermore National Laboratory, 7000 East Ave., Livermore, CA 94551, USA}

\author{T.P.~Biesiadzinski} 
\affiliation{Case Western Reserve University, Department of Physics, 10900 Euclid Ave, Cleveland, OH 44106, USA}
\affiliation{SLAC National Accelerator Laboratory, 2575 Sand Hill Road, Menlo Park, CA 94205, USA}
\affiliation{Kavli Institute for Particle Astrophysics and Cosmology, Stanford University, 452 Lomita Mall, Stanford, CA 94309, USA}

% ...

\collaboration{LUX Collaboration}

\date{\today}

\begin{abstract}
This work presents an analysis of monoenergetic electronic recoil peaks in the dark-matter-search and calibration data from the first underground science run of the Large Underground Xenon (LUX) detector. Liquid xenon charge and light yields for electronic recoil energies between 5.2 and 661.7 keV are measured, as well as the energy resolution for the LUX detector at those same energies. Additionally, there is an interpretation of existing measurements and descriptions of electron-ion recombination fluctuations in liquid xenon as limiting cases of a more general liquid xenon recombination fluctuation model. Measurements of the standard deviation of these fluctuations at monoenergetic electronic recoil peaks exhibit a linear dependence on the number of ions for energy deposits up to 661.7 keV, consistent with previous LUX measurements between 2-16~keV with $^3$H. We highlight similarities in liquid xenon recombination for electronic and nuclear recoils with a comparison of recombination fluctuations measured with low-energy calibration data.
\end{abstract}

\pacs{Valid PACS appear here}
\maketitle

\section{\label{Intro}The LUX Detector}

The Large Underground Xenon (LUX) detector is a two-phase (liquid/gas) xenon time-projection chamber (TPC) designed to detect weakly-interacting massive particles (WIMPs), a favored dark matter candidate \cite{NIM}. LUX has produced world-leading exclusion limits for spin-independent and spin-dependent WIMP-nucleon scattering cross-sections \cite{PRL,reanalysis,run4,spinDep}. The detector uses a dodecagonal active volume with 251~kg of liquid xenon (LXe), bounded in $z$ by cathode and gate wire grids (48.3~cm apart) and in $(x, y)$ by 12 PTFE panels (47.3~cm face-to-face) \cite{php}. The active volume is monitored by 122 photomultiplier tubes (PMTs) that are divided evenly between top and bottom arrays. Energy depositions produce prompt scintillation light (S1) and delayed electroluminescence light (S2) created by drifting liberated ionization electrons via an applied electric field from the interaction site to the liquid surface. An even higher field is applied between the gate and anode grids at the surface, and the electrons are extracted into gaseous xenon to produce the S2. The time, $t_d$, between the S1 and S2 signals defines the depth of the interaction and the $(x_{S2}, y_{S2})$-position is reconstructed from the S2 hit pattern in the top PMT array \cite{mercury}. Further technical detail on the LUX detector can be found in \cite{NIM}.

The ratio of free charge to scintillation light, typically expressed as $\log_{10}\left(\text{S2/S1}\right)$, is used to distinguish electronic recoils (ER) and nuclear recoils (NR) produced by incoming particles interacting with xenon atoms. Discrimination between ER and NR events makes LXe TPC detectors viable dark matter discovery experiments. The underlying microphysics of these recoils is an area of active and robust modeling, most notably by the Noble Element Simulation Technique (NEST) \cite{NEST2011}. Critical to these models is the measurement of light and charge yields for xenon at a wide range of energies. LUX has previously measured ER absolute light and charge yields down to 1.3 keV with a novel \emph{in situ} $^3$H calibration \cite{CH3T}. The complementary measurements of the light and charge yields at higher energies ($>$10 keV) beyond the WIMP search region follow here. Measurement and calibration of the LXe response beyond the WIMP-search energy range is relevant for any potential Compton imaging applications, neutrinoless double beta decay searches, and the understanding of backgrounds that extend into the search regions for WIMPs and other potential dark matter candidates. Additionally, these measurements constrain theoretical models for charge and light production in liquid xenon, notably the transition region between the Thomas-Imel (ER energies $\lesssim$10~keV) and Doke ($\gtrsim$10~keV) recombination models \cite{TIB, DokeRecomb}.

\begin{figure}
	\includegraphics[scale=0.38]{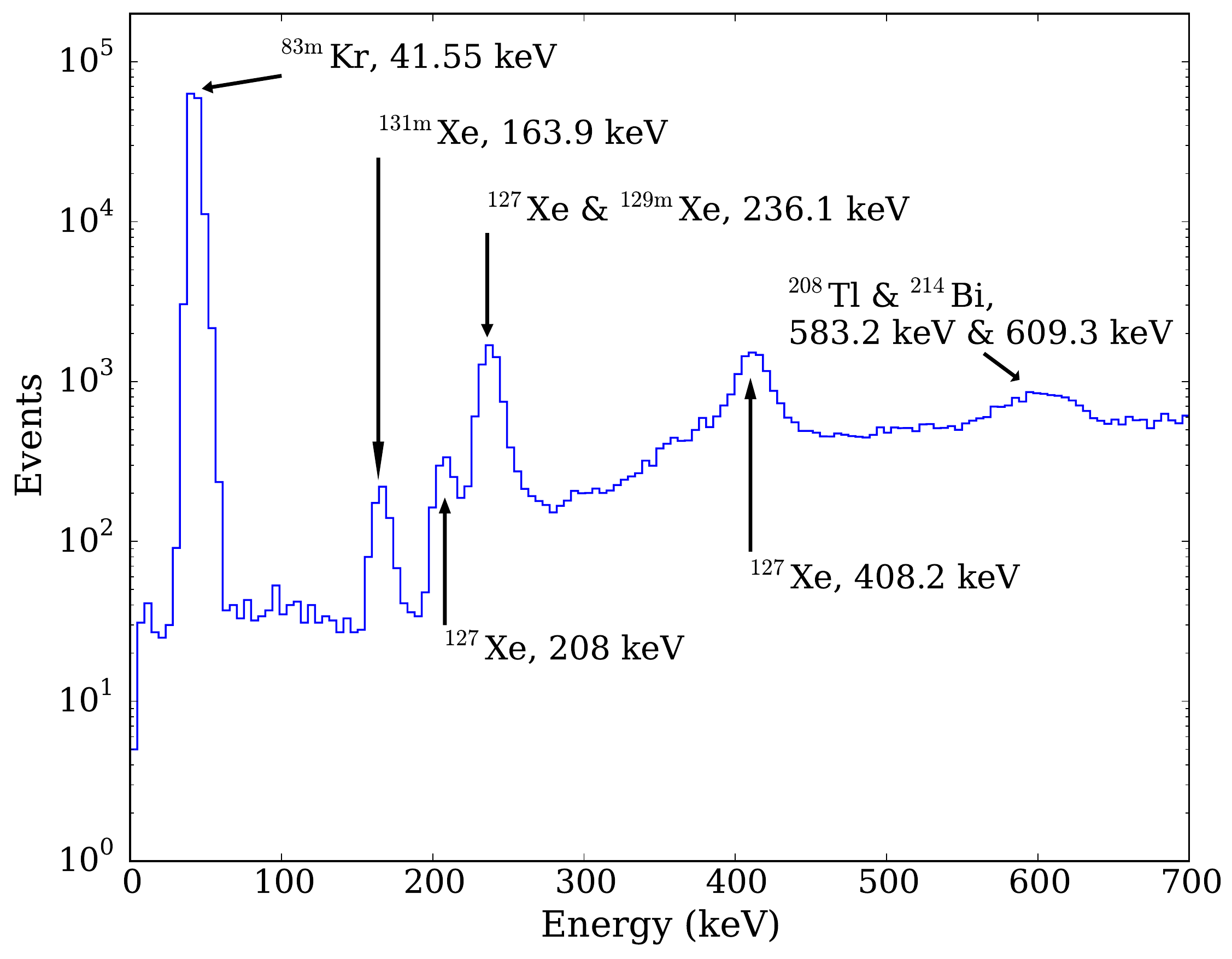}
    \caption{\label{especWS} Single-scatter events identified in the LUX 2013 WIMP-search data. The labels indicate the source isotopes and their energies. Only radial and drift-time fiducial cuts ($r_{S2}<20$~cm; 38~$\mu\textrm{s}<t_d<305$~$\mu\textrm{s}$) have been applied to make this plot; additional cuts are applied to maximize signal-to-background for each peak individually in the following measurements.}
\end{figure}

\section{\label{Microphys}Energy reconstruction and signal yields}

Particle interactions in liquid xenon excite atoms (forming excitons), create electron-ion pairs, and produce atomic motion (heat). Energy in the first two channels yield photons and electrons, \textit{i.e.} detectable quanta, while the amount of energy in the form of heat is negligible for electronic recoils. Therefore, energy depositions can be described with
\begin{equation}
\begin{aligned}
  E &= f W (n_{ex} + n_i) \\
  &= f W (1 + \frac{n_{ex}}{n_{i}}) n_i , 
\end{aligned}
\end{equation}
where $E$ is the energy, and $n_{ex}$ and $n_i$ are the numbers of excitons and electron-ion pairs, respectively \cite{Platz}. $W$ is the average energy needed to produce a single excited or ionized atom and its value is $W = 13.7\pm0.2$ eV \cite{Dahl}. The quenching factor, $f$, is negligible for electronic recoils and thus $f \equiv 1$ in this paper; LUX NR ($f\neq1$) measurements can be found in \cite{DDpaper} and in a brief discussion in Sec. \ref{NRfluct}. The ratio of excitons to ions is constant for ER interactions, $n_{ex}/n_{i} = 0.2$ \cite{TIB,exion1,exion2}. Each exciton de-excites, emitting a 178-nm photon \cite{Mock,XePhys1,XePhys2}. A fraction of the initial electron-ion pairs, $r$, recombine and form additional excitons. Electron-ion recombination is a fundamental property of liquid xenon that depends on the fluid density, applied electric field, and particle energy \cite{Conti,Dahl,exion2,CH3T}. The measurements presented here were made with an average drift field of 180~V/cm as in \cite{reanalysis}. There is a slight degeneracy between $n_{ex}/n_{i}$ and $r$, particularly if $n_{ex}/n_{i}$ exhibits an energy dependence. $n_{ex}/n_{i}=0.2$ is consistent within uncertainties of the most recent measurements \cite{Kaixuan}, and it is held constant for simplicity.
\begin{figure}
	\includegraphics[scale=0.39]{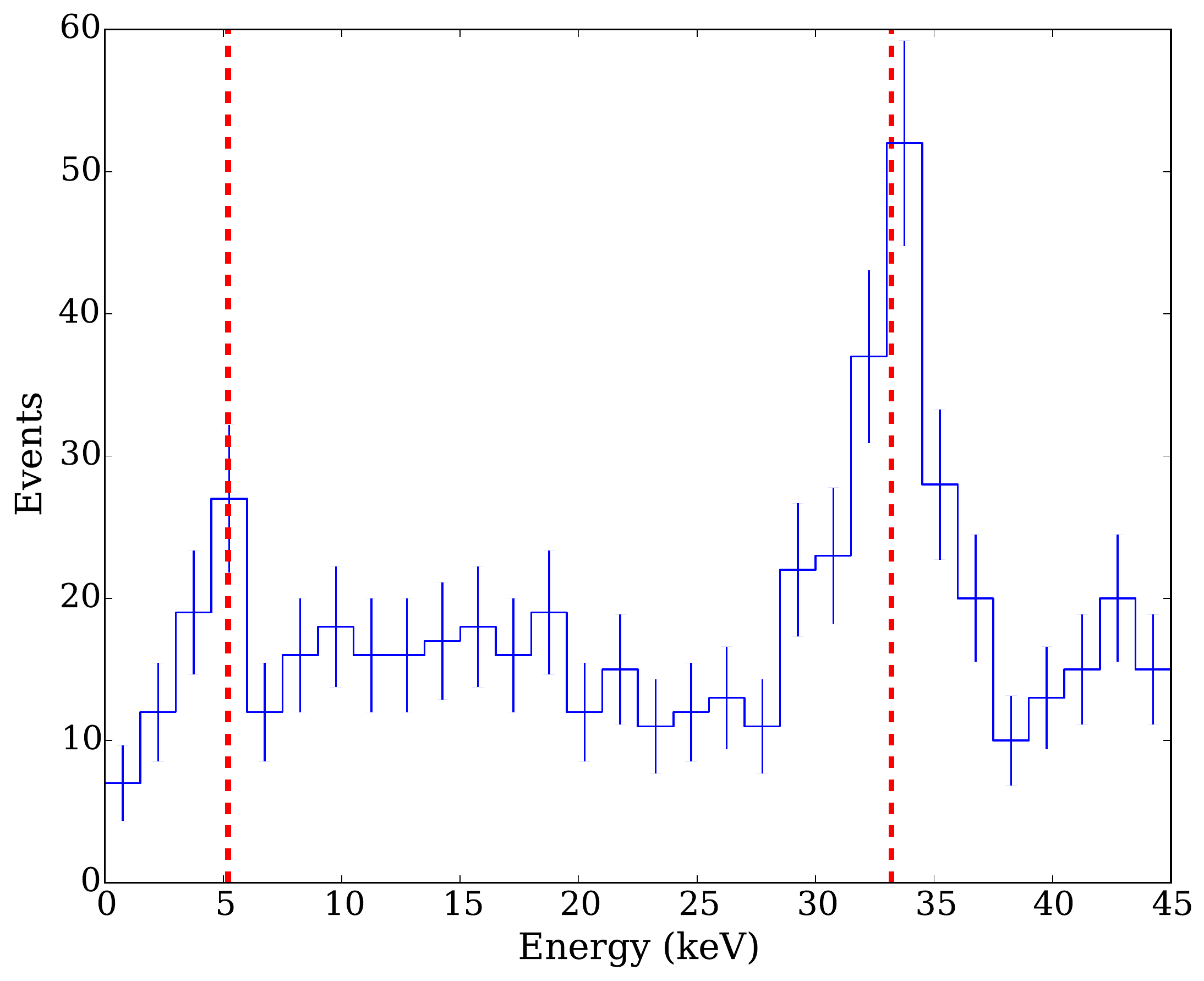}
    \caption{\label{especWSlow} Single-scatter events satisfying radial and drift-time fiducial cuts ($r_{S2}<20$~cm; 38~$\mu\textrm{s}<t_d<305$~$\mu\textrm{s}$) prior to May 12 in the LUX 2013 WIMP-search data. The error bars represent statistical uncertainty. The dashed red lines indicate the true peak energies of the 5.2 and 33.2 keV, $L$- and $K$-shell, electron captures (EC) of $^{127}$Xe. The 33.2 keV $K$-shell peak emerges clearly when data collected in the 24 hours following each $^{83\textrm{m}}$Kr injection are excluded. $^{127}$Xe was cosmogenically produced while in storage above ground, and it began decaying away once the LUX xenon was moved underground. These EC sources became neglibly weak by the end of the WIMP search data acquisition.}
\end{figure}
\begin{figure}
	\includegraphics[scale=0.38]{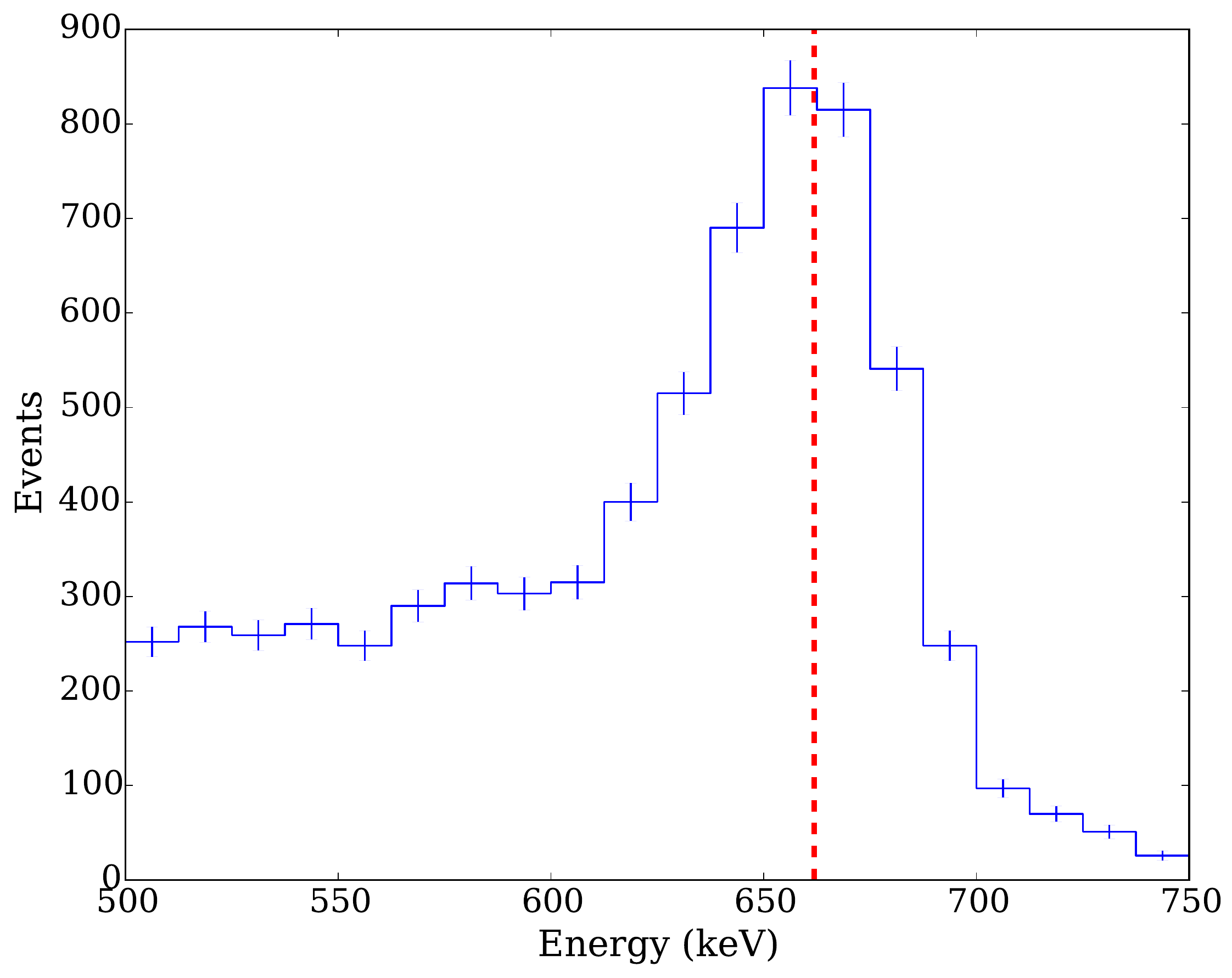}
    \caption{\label{especCs} Fiducial single-scatter events ($r_{S2}<20$~cm; 38~$\mu\textrm{s}<t_d<305$~$\mu\textrm{s}$) from an August 2013 $^{137}$Cs calibration of the LUX detector. The dashed red line indicates the true energy of the photopeak at 661.7 keV.}
\end{figure}

In practice, the directly measurable quantities are the de-excitation photons (from initial and recombined excitons) and the electrons that escape recombination. They are expressed as
\begin{equation}
n_{\gamma} = (\frac{n_{ex}}{n_{i}} + r)n_i
\end{equation}
and
\begin{equation}
n_e = (1-r)n_i,
\end{equation}
and these relate directly to the S1 and S2 signals recorded in LUX. In terms of S1 and S2, we rewrite the expression for energy
\begin{equation}
\begin{aligned}
  E &= W\left(n_{\gamma} + n_e\right) \\
  &= W\left(\frac{S1}{g_1} + \frac{S2}{g_2}\right) , 
\end{aligned}
\end{equation}
where $S1$ and $S2$ in units of detected photons (phd) are pulse sizes corrected for geometrical effects and electron lifetime in LXe \cite{reanalysis}. The detector gains, $g_1$ and $g_2$, are in units of phd/quantum. $g_1$ represents the overall photon detection efficiency for prompt scintillation in the liquid and is the product of the LUX average light collection efficiency and the average PMT quantum efficiency. $g_2$ is the corresponding quantity for S2 light, consisting of the product of the electron extraction efficiency (from liquid to gas) and the average single electron pulse size in phd. For the data analyzed in this work, these detector-specific quantities have been measured to be $g_1 = 0.117 \pm 0.003$ phd/photon and $g_2 = 12.1 \pm 0.8$ phd/electron, with an electron extraction efficiency of $49\%\pm3\%$ \cite{reanalysis}. Used in Eq.~4, they allow for the energy reconstruction of ER interactions observed in the LUX detector. Explicitly, the light ($L_y$) and charge ($Q_y$) yields are defined as
\begin{equation}
L_y = \langle n_\gamma \rangle / E
\end{equation}
and
\begin{equation}
Q_y = \langle n_e \rangle / E.
\end{equation}

\section{\label{DataSelection}Data Selection}

The energy spectrum of single-scatter events acquired during the LUX 2013 WIMP search, shown in Figs. 1 and 2, includes peaks from the $^{127}$Xe $L$-shell electron capture at 5.2~keV to the 609~keV gamma emitted following $^{214}$Bi $\beta$-decay. There is a large contribution at 41.6 keV from residual $^{83\text{m}}$Kr, an internal calibration source injected regularly during the acquisition \cite{Kastens:JINST,Manalaysay:2009yq}. Figure \ref{especCs} shows part of the Compton plateau and the 661.7 keV photopeak from $^{137}$Cs calibrations. All energies and sources of relevant peaks are listed in Table \ref{Etable}.

\begin{table}
\begin{ruledtabular}
\begin{tabular}{ccl}
    Energy (keV) & Source & Decay\\
    \hline
    5.2 & $^{127}$Xe & $L$-shell electron capture (EC) \\
    33.2 & $^{127}$Xe & $K$-shell EC \\
    41.55 & $^{83\textrm{m}}$Kr & 32.1 + 9.4 keV conversion electrons \\
    163.9 & $^{131\textrm{m}}$Xe & 163.9 keV gamma \\
    208.1 & $^{127}$Xe & $L$-shell EC + $^{127}$I 202.9 keV gamma \\
    236.1 & $^{127}$Xe & $K$-shell EC + $^{127}$I 202.9 keV gamma \\
    236.1 & $^{129\textrm{m}}$Xe & 196.6 + 39.6 keV gammas \\
    408.2 & $^{127}$Xe & $K$-shell EC + $^{127}$I 375 keV gamma \\
    583.2 & $^{208}$Tl & $\beta$-decay + $^{208}$Pb 583.2 keV gamma \\
    609.3 & $^{214}$Bi & $\beta$-decay + $^{214}$Po 609.3 keV gamma \\
    661.7 & $^{137}$Cs & $\beta$-decay + $^{137}$Ba 661.7 keV gamma \\
\end{tabular}
\end{ruledtabular}
\caption{\label{Etable} The energies and details of each peak in the ER energy spectrum. The $^{129\textrm{m}}$Xe decay and one of the $^{127}$Xe processes completely overlap at 236.1~keV. There are additional decay schemes that are in, or near, the 208.1 and 408.2~keV peaks with lower rates ($<$10\% relative to these modes).}
\end{table}

Cosmogenically activated isotopes $^{127}$Xe, $^{129 \textrm{m}}$Xe, and $^{131\textrm{m}}$Xe decay with half lives of 36.3, 8.9, and 11.8 days, respectively. As short-lived intrinsic sources, their signal is maximized relative to Compton backgrounds by including data from only the first 20 days of the WIMP search for all Xe activation peaks and by applying an $r_{S2}=18$~cm fiducial cut as in \cite{PRL} for peaks at 163.9, 208.1, 236.1, and 408.2 keV. $^{127}$Xe is responsible for five of the peaks in this study (Figs.~\ref{especWS} and \ref{especWSlow}). Its decay is characterized by an electron capture followed immediately by the de-excitation of $^{127}$I. A dedicated study of $^{127}$Xe decay using the LUX detector is forthcoming in \cite{DQ127}. The peaks for the $L$- and $K$-shell $^{127}$Xe electron captures (5.2 and 33.2 keV) are fitted using events with $r_{S2}<20$~cm because these events occur near the periphery where the $^{127}$I gamma can escape without depositing energy in the active region. Additionally, for the $^{127}$Xe $K$-shell peak, we exclude data occurring within 24 hours of $^{83\textrm{m}}$Kr calibration injections to avoid contamination from its 41.55~keV decay. $^{83\textrm{m}}$Kr has a 1.85-hour half life, and this cut removes $>$$99.99\%$ of all events with energies reconstructed between 40 and 43~keV.

The cuts for detector stability and event/pulse quality in this analysis are the same cuts used in the WIMP-search analysis \cite{LUX:PRD}. Detector stability cuts exclude data from periods with excursions from normal detector conditions and times immediately following power outages and circulation stoppages (0.8\% reduction in livetime). An event quality cut limits the combined waveform area outside of S1 and S2 pulses within the same 1 ms event window. It excludes events that have additional pulse area that is more than $10\%$ of the combined pulse areas of S1 and S2 in the waveform, which leads to a 1\% reduction in livetime. This cut removes events with large numbers of spurious single photoelectrons or extracted electrons.

\section{\label{Yields}Analysis of LUX Data}

\begin{figure}
	\includegraphics[scale=1]{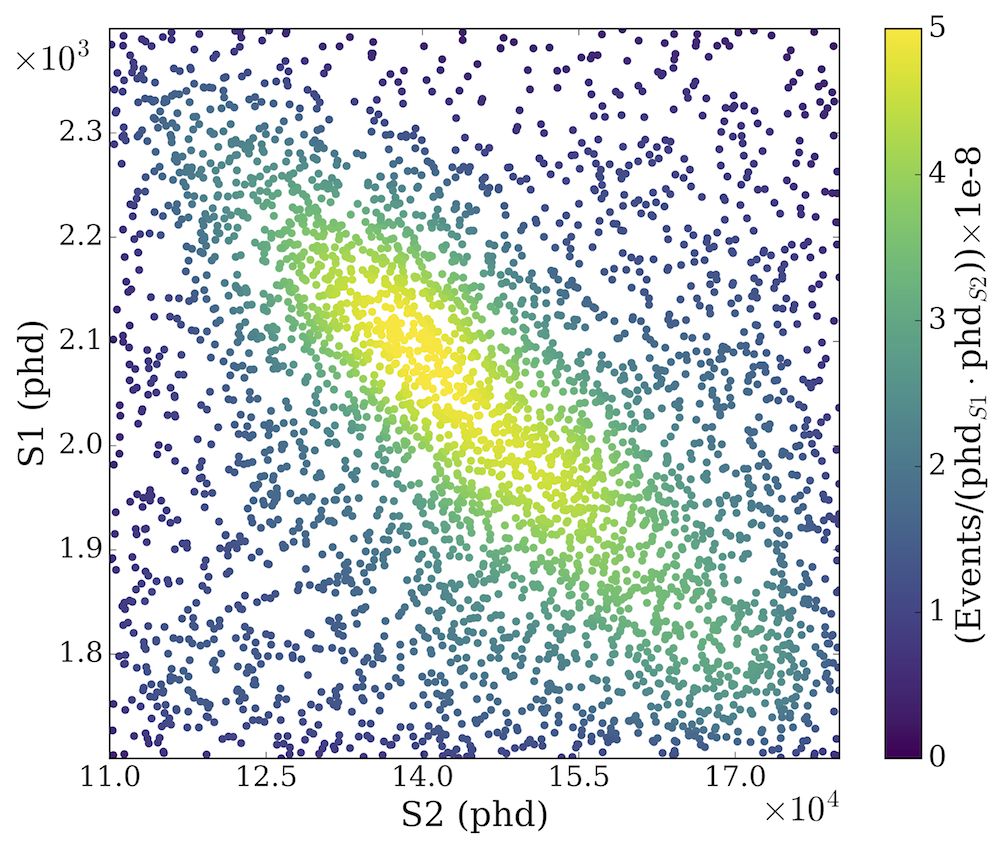}
    \caption{\label{s1s2plot} The S1 and S2 corrected pulse areas from events including 410-keV $^{127}$Xe decays.}
\end{figure}
\subsection{\label{analysisSignal}Signal Yields}

\begin{figure}
	\includegraphics[scale=0.49]{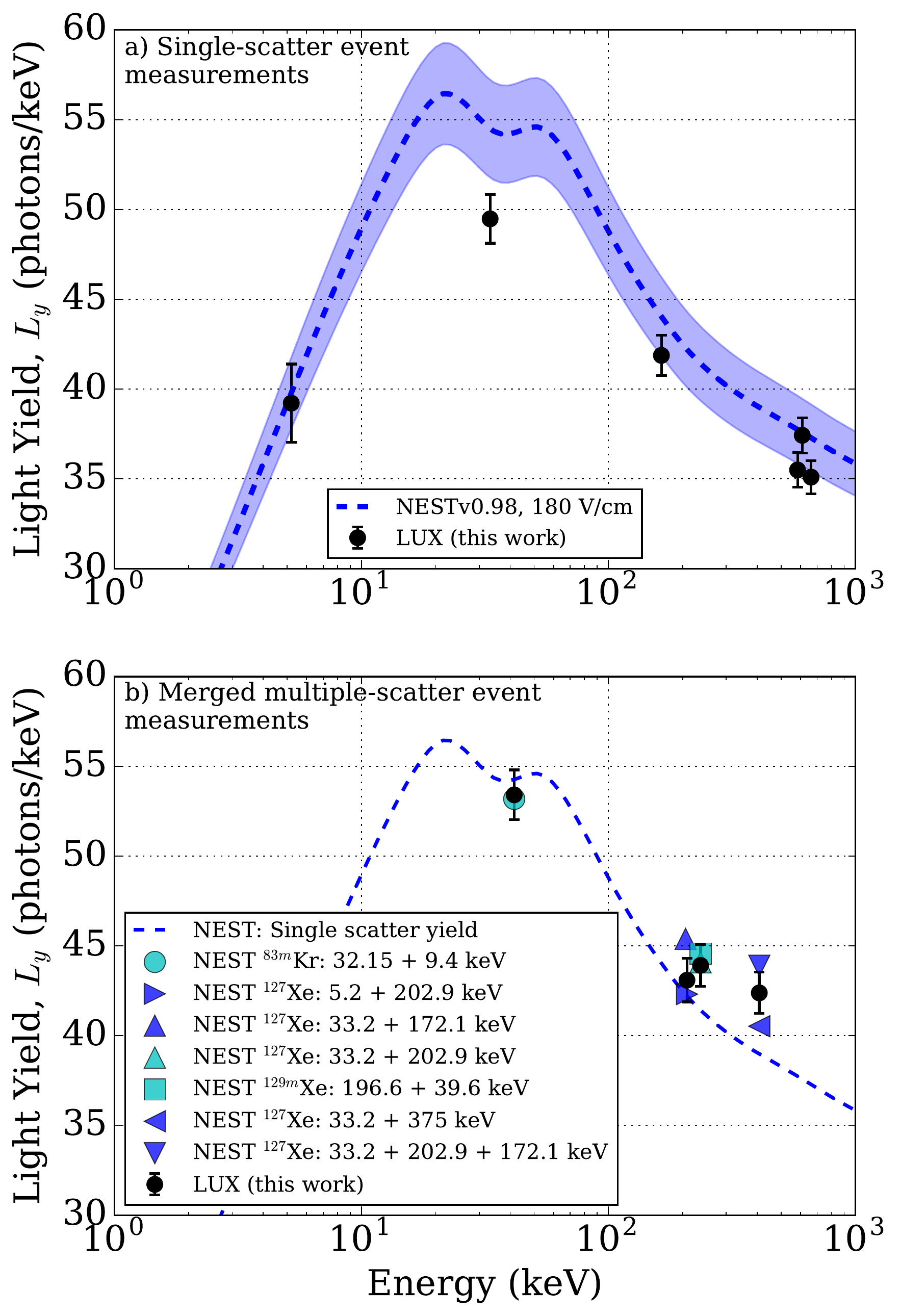}
    \caption{\label{lyplot} The measured light yield at peak energies in the LUX ER energy spectrum for (a) single- and (b) multiple-site energy depositions. In (a), the dashed blue line is the mean response predicted by NEST for an applied field of 180~V/cm, and the shaded blue region shows its 5$\%$ uncertainty. In (b), the NEST predictions for multi-component decays of $^{83\textrm{m}}$Kr, $^{127}$Xe, and $^{129\textrm{m}}$Xe are made by summing the mean photons expected from the constituent scatters and dividing by the total energy. The light yield of 33.2~keV $^{127}$Xe is lower than the NEST prediction, where the recombination models transition.}
\end{figure}

\begin{figure}
	\includegraphics[scale=0.49]{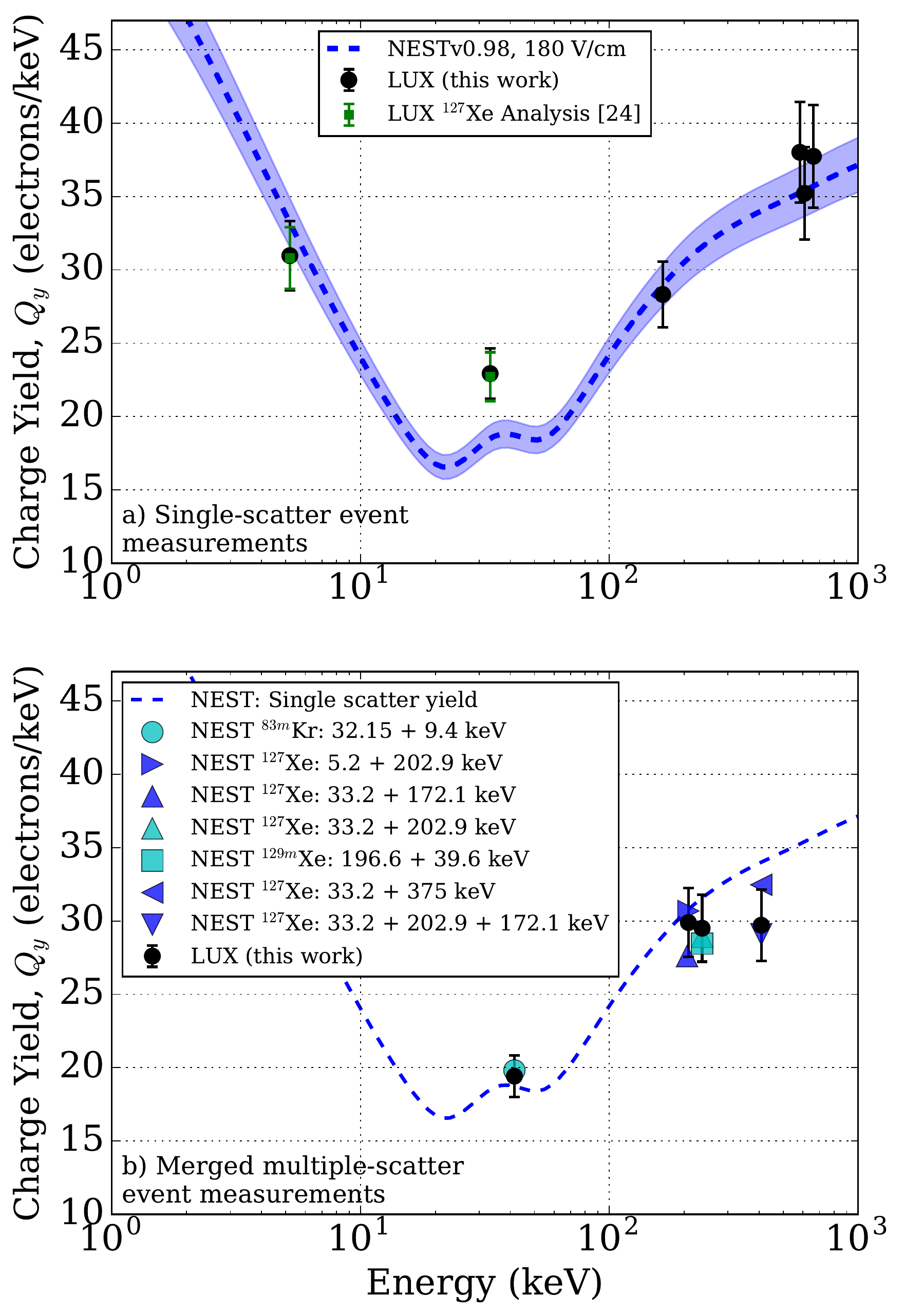}
    \caption{\label{qyplot} The measured charge yield at peak energies in the LUX ER energy spectrum for (a) single- and (b) multiple-site energy depositions. In (a), the dashed blue line is the mean response predicted by NEST for an applied field of 180~V/cm, and the shaded blue region shows its 5$\%$ uncertainty. Charge yields measured at 33.2 and 5.2~keV in a dedicated two-S2 $^{127}$Xe LUX analysis (green) \cite{DQ127} agree with these one-S2 measurements (black). In (b), the NEST predictions for multi-component decays of $^{83\textrm{m}}$Kr, $^{127}$Xe, and $^{129\textrm{m}}$Xe are made by summing the mean electrons expected from the constituent scatters and dividing by the total energy.}
\end{figure}

Each monoenergetic source generates a fixed mean amount of light and charge. Monoenergetic signals appear as elliptical overdensities in ($S1,S2$)-space as plotted in Fig. \ref{s1s2plot}. The major axis of the ellipse follows a line of constant energy, with the length of that axis dictated by recombination fluctuations. Additional spread in the S1 and S2 response for a monoenergetic source comes from the finite detector resolution in the respective channels. Fits for the mean S1 and S2 response at each energy are made with data within $2\sigma$ of the mean reconstructed energy. Measurements of the light and charge yields, shown in Figs. \ref{lyplot} and \ref{qyplot}, follow directly from Gaussian fits for the mean S1 and S2 as described in Sec.~\ref{Microphys}.
Figs.~\ref{lyplot} and \ref{qyplot} show comparisons of these LUX measurements with the most recently published NEST models for light and charge yields at 180~V/cm. The upper (a) panels show the measured signal yields of the single-site energy depositions along with the functional form of the NEST model plotted for comparison. The dashed blue line is the mean response predicted by NEST for an applied field of 180~V/cm, and the shaded blue region shows its 5$\%$ uncertainty \cite{NEST2011}. The uncertainty in the NEST model comes from the dispersion of the world's data and interpolating to the LUX-specific applied drift field. The lower (b) panels show the measured signal yields from multiple-site energy depositions where the light and charge quanta from the lower-energy constituent decays are merged. The NEST yields from the specific energies of the possible decay modes within each monoenergetic peak are summed and plotted with the LUX measurements for comparison.

\subsection{Mean Recombination}
\begin{figure}
	\includegraphics[scale=0.483]{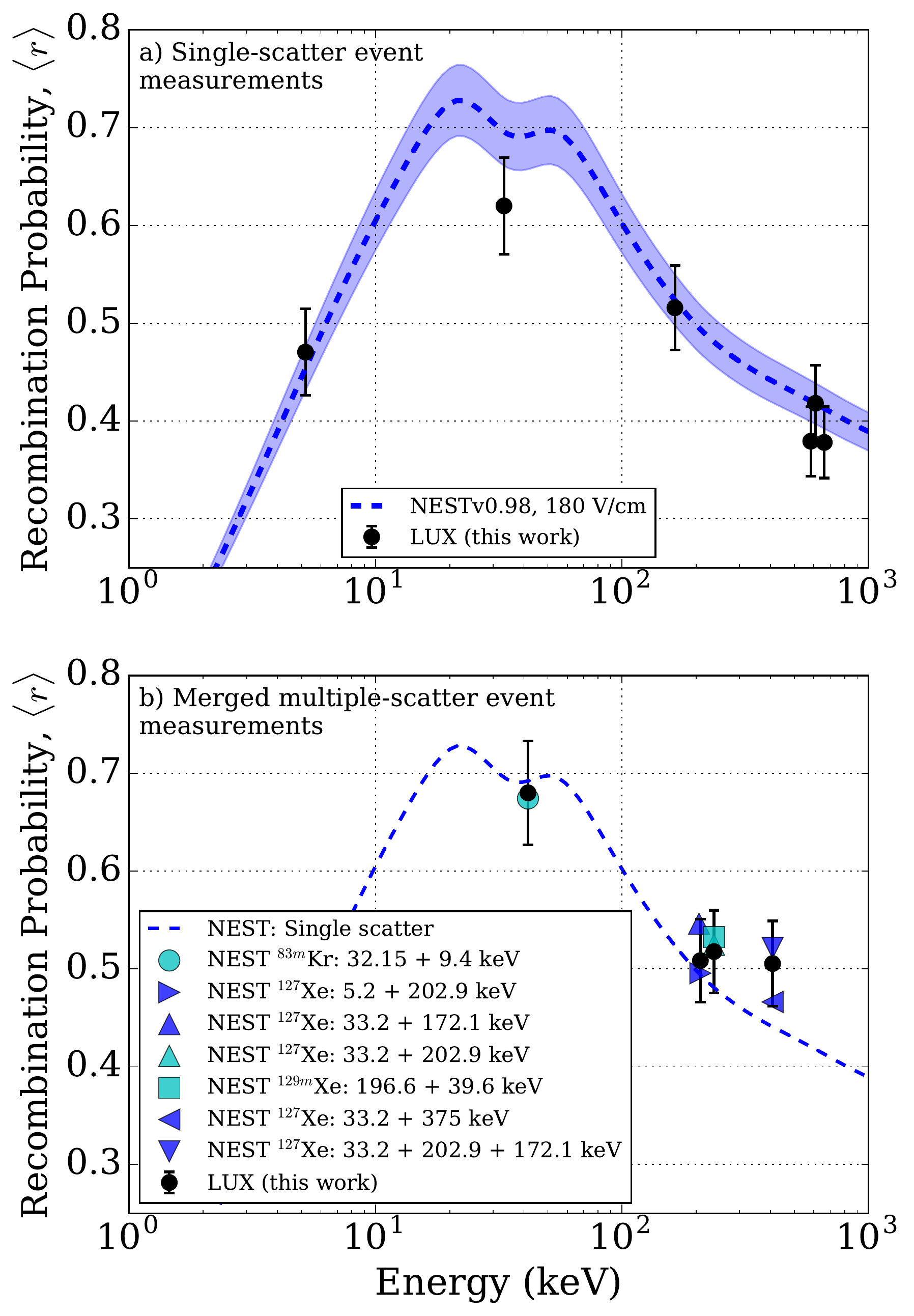}
    \caption{\label{recombplot} The recombination probability calculated at peak energies in the LUX ER energy spectrum for (a) single- and (b) multiple-site energy depositions. In (a), the dashed blue line is the mean recombination predicted by NEST for an applied field of 180~V/cm, and the shaded blue region shows its 5$\%$ uncertainty.}
\end{figure}

From Eqs. 1-4 in Sec.~\ref{Microphys}, one can obtain the mean recombination probability $\langle r \rangle$
\begin{equation}
\langle r \rangle = \frac{ \langle n_{\gamma} \rangle /\langle n_{e} \rangle - {n_{ex}}/{n_{i}}}{\langle n_{\gamma} \rangle /\langle n_{e} \rangle + 1} ,
\end{equation}
where $\langle n_{\gamma} \rangle /\langle n_{e} \rangle \equiv L_y / Q_y$ is directly proportional to the measured mean S2/S1. The LUX values for $\langle r \rangle$ are shown in Fig.~\ref{recombplot}, with single- and multiple-site energy depositions plotted separately for comparison with NEST as in Sec.~\ref{analysisSignal}.


\subsection{Energy Resolution}
\begin{figure}
	\includegraphics[scale=0.44]{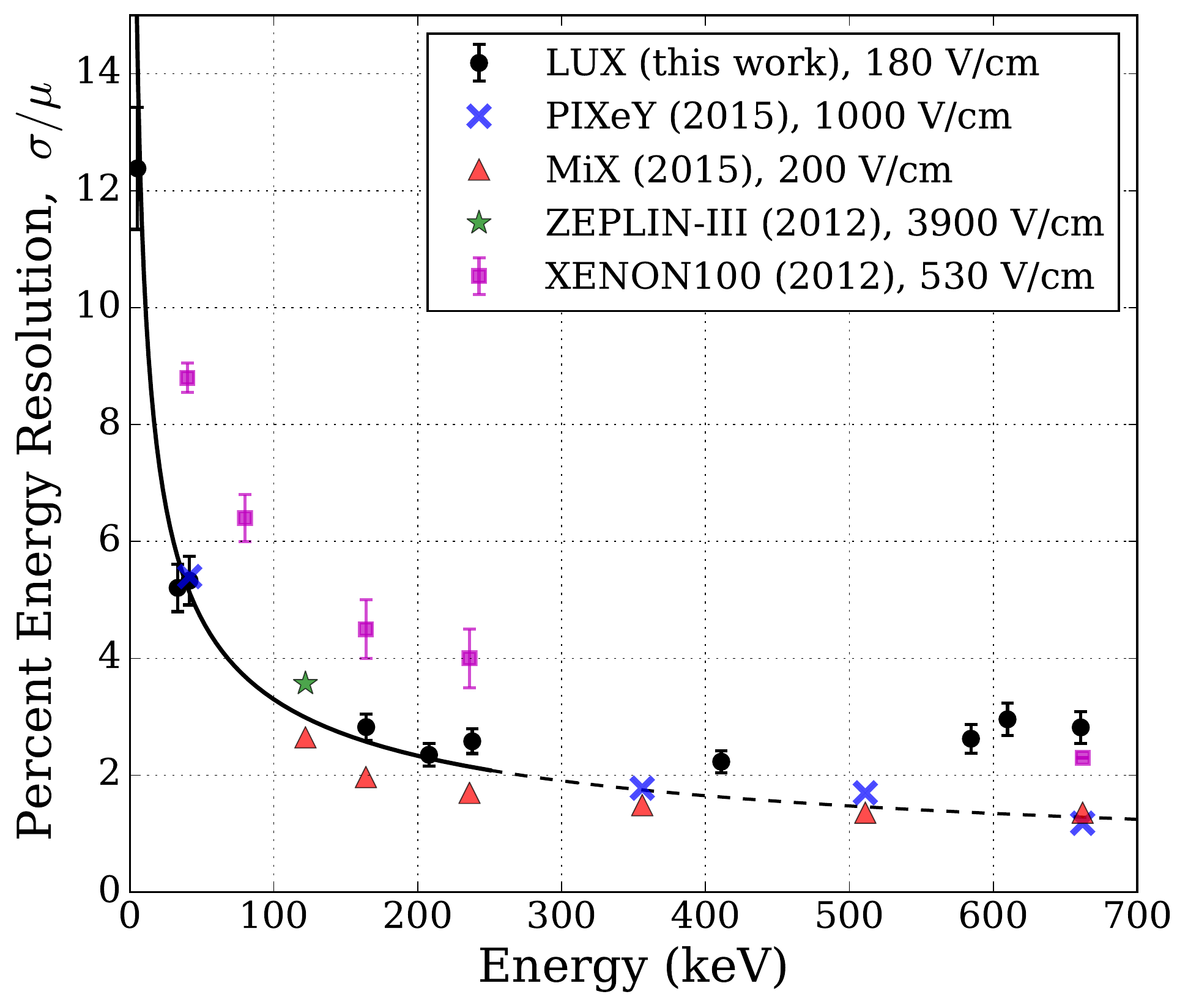}
    \caption{\label{fig:eresplot} The measured energy resolution at known energy peaks in the LUX ER backgrounds. The detector is optimized for low energy sensitivity, and variable amounts of PMT saturation and single-electron contributions affect S2 pulses and hamper the energy resolution at high energy, as discussed in the text. Data from the PIXeY (blue x; \cite{PIXeYthesis,PIXeY}), MiX (red triangle; \cite{MiX}), ZEPLIN-III (green star; \cite{Akimov}), and XENON100 (magenta square; \cite{Xenon100}) are shown for comparison.}
\end{figure}

In measuring $L_y$ and $Q_y$ from monoenergetic sources, one also easily measures the energy resolution. These measurements are shown in Fig.~\ref{fig:eresplot}. An empirical fit of the form $a/\sqrt{E}$ to the LUX measurements made at the six lowest energies in Fig.~\ref{fig:eresplot} yields $a=(0.33 \pm 0.01 \text{ keV}^{1/2})$, and it is plotted in solid black over the fit range and dashed where it is extrapolated. The energy resolution observed above $\gtrsim240$~keV is worse than the expected resolution from a fit with only a stochastic $1/\sqrt{E}$ term of the values from $<240$~keV monoenergetic sources.

\subsection{Discussion of results}

These results agree well with the expected yields and mean recombination predicted by NEST, except the measurements made at 33.2 keV. The LUX measurement is far from threshold and of a low enough energy to be free from the soon-to-be-discussed S2 systematics. Disagreement with NEST is not completely unexpected: that particular energy is a difficult one to model because the accepted understanding of LXe recombination transitions from a spherically distributed cloud of electron-ion pairs below $\sim$10 keV \cite{TIB} to a track-like structure of electron-ion pairs above that energy \cite{DokeRecomb}. The charge yield of the same 33.2~keV $K$-shell energy measured with a separate multiple-scatter analysis of LUX $^{127}$Xe data produced the same result with similar levels of uncertainty \cite{DQ127}. The LUX energy resolution at energies below 250~keV compares favorably with previous measurements in large LXe TPCs \cite{Xenon100} and is comparable to the resolution achieved by ZEPLIN-III \cite{Akimov} and by much smaller detectors \cite{MiX,PIXeYthesis,PIXeY}. Tables \ref{tab:lyqy} and \ref{tab:epeaks} near the end of this article list the LUX values and uncertainties plotted in Figs.~\ref{lyplot}-\ref{fig:eresplot}. Larger systematic uncertainties in the charge yield and recombination measurements (and poorer energy resolution) above 250~keV stem from the following S2 effects in LUX.

First, the amount of S2 electroluminescence at energies greater than 500~keV is enough to exceed the maximum of the data acquisition (DAQ) digitization range for one or more PMTs in top array. The amount of saturation depends on the $(x, y)$ position at which the extracted electrons emerge from the liquid beneath the top array of PMTs, which broadens the spectrum of S2 pulse areas and skews it towards lower pulse areas. This is results in an additional $6\%$ bias in the S2 measurements at energies above 500~keV, which is measured by comparing the ratio of S2 pulse area observed in the bottom PMT array to the area in both arrays ($S2_b$/$S2$) to the same ratio of pulse areas for single extracted electrons. It is also observed that S2 pulses from high energy events have tails of electroluminescence created by extracted electrons trailing the primary pulse. A variable amount of this ``electron tail" is folded into the S2 pulse area, introducing an area-dependent uncertainty in the S2 measurement. The origin of these electrons has been studied in previous LXe TPCs with two main sources identified: the delayed extraction of electrons from previous energy deposits and the production and the extraction of additional electrons from optical feedback due to the quantum efficiency of the electrode grids and from photoionization of impurities in the LXe bulk \cite{zeplinSantos}. To quantify the additional pulse area from the electron tail, we compare the total area found by the pulse finder (S2 pulse and possible electron tail) to the area calculated from a Gaussian fit to the primary pulse. The Gaussian model is an approximation for an idealized S2 pulse shape without a single-electron tail. By this method we calculate a $6.1\%$ systematic bias in the charge yield at 661.7~keV, scaling linearly to $3.1\%$ at 163.9~keV. For S2 pulses with $\lesssim$10$^4$~phd this effect is sub-dominant to uncertainties in $g_1$ and $g_2$. The mean DAQ saturation and the mean single electrons tail contributions are nearly equal and opposite effects. The combined effect minimally affects the central values of the signal yields and mean recombination measurements, but broadens the spectrum of S2 pulse areas and impacts the energy resolution for the four highest energy peaks considered in this work.

Finally, events with sub-cathode scatters were addressed. Referred to as ``gamma-X," this is a multi-site interaction where the gamma scatters at least once below the cathode wire grid and only once above it. When this happens, the detector collects scintillation light from all interaction vertices but charge from only the interaction above the cathode, where the electric field drifts electrons upwards to the gas layer for S2 production. Gamma-X events are misclassified in the data processing as single-site interactions with a larger S1 and smaller S2 relative to events of the same reconstructed energy. These events are more common at high energies where the gamma from radioactive decay within detector materials has sufficient energy to travel several centimeters into and between the fiducial and sub-cathode volumes. This pathology is excluded from the analysis dataset by selecting events from a smaller fiducial volume further from the cathode plane, and also requiring a minimum S2 size ($\textrm{S2} > \left(\langle\textrm{S2}\rangle - 2\sigma_{\textrm{S2}}\right)$ within each monoenergetic peak). These additional cuts reduced gamma-X contamination to less than $1\%$ of its initial level measured in the distribution of S2 areas from events within 3$\sigma$ of each peak in reconstructed energy.

The net effect of the DAQ saturation, single-electron tail fraction, and remaining gamma-X (after additional S2 area cuts) makes the energy resolution 2.0 times worse than the expectation from the stochastic term alone at the peaks below 240 keV. With optimized PMT DAQ settings and electron extraction efficiency at or near unity, two-phase Xe TPC detectors have demonstrated $\sigma/\mu \leq 1\%$ capability at 2.6 MeV, the energy regime relevant for $0\nu\beta\beta$ searches with $^{136}$Xe \cite{MiX,PIXeY}. Some signal fluctuations are ultimately unavoidable, however, due to recombination fluctuations in the LXe itself, as discussed in the next section.

\begin{figure}
	\includegraphics[scale=0.412]{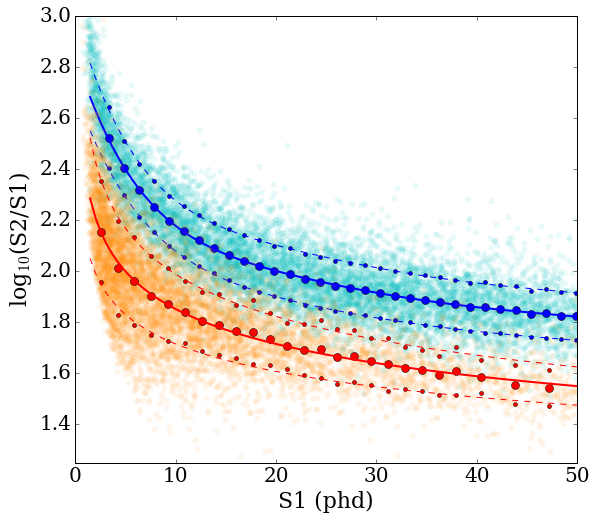}
    \caption{\label{fig:bandsplot} The ER and NR calibration data (cyan and orange, respectively) form characteristic recoil bands. Large filled circles show the fitted band Gaussian mean and small filled circles indicate the fitted Gaussian $\pm$1$\sigma$. Power law fits to the means and $\pm$1$\sigma$ are shown with solid and dashed lines.}
\end{figure}

\section{Recombination fluctuation models and analysis}\label{Flucts}

It has been known for decades that fluctuations in electron recombination in liquid xenon exhibit a variance in excess of the expectation for a binomial distribution \cite{TIB}. In the context of dark matter search experiments, this variance manifests itself in the width (in $\log_{10}$(S2/S1)) of the electronic recoil band, shown in Fig.~\ref{fig:bandsplot}. To a high but imperfect degree, this band appears Gaussian in slices of S1 \cite{CH3T}.

One approach to analyzing the data is to
\begin{enumerate}[label=(\alph*)]
\item subtract the (calculable) instrumental fluctuations, and fit the remaining recombination fluctuations with a Gaussian, characterized by $\sigma _r$. This approach was followed in \cite{CH3T, Dobi} and results in the somewhat surprising observation that $\sigma _r$ grows linearly with the number of ions created by the interaction, rather than scaling as $\sqrt{n_i}$ as would be expected.
\end{enumerate}

A slightly different approach is taken by the NEST model, which is described in detail in \cite{NEST2014}. The key difference in the present context is that NEST 
\begin{enumerate}[label=(\alph*)]
\setcounter{enumi}{1}
\item accounts for all fluctuations using a modified Poisson distribution. A Poisson distribution is chosen to avoid the computational expense of a binomial distribution. The modification assigns the Poisson distribution's average number of quanta (expressed as $\lambda$) from a Gaussian distribution, creating the desired observed width of fluctuations while respecting physical constraints (integer quanta with $n_i\geq0$) \cite{NEST2014}. The width of this Gaussian distribution is determined empirically from calibration data.
\end{enumerate}

Both of these approaches are explored in the present work, so it is worth pointing out that they are essentially limiting cases of the same general picture, discussed in more detail below.

\subsection{General Picture}
Approach (a) and (b) are approximations to a more general description. In the limit of isolated electron-ion pairs, one might reasonably expect recombination to be a binomial process governed by an escape probability $p \equiv 1-r$, so that the number of measured electrons is
\begin{equation}
n_e = \binom{n_i}{p} .
\end{equation}
At rather low electronic recoil energies $E\lesssim10$~keV, it can be shown that the Thomas-Imel model \cite{TIB} reproduces the central value of this probability
\begin{equation} \label{eq:TI}
p = \frac{1}{\xi} \log\left(1+\xi\right) ,
\end{equation}
where $\xi$ is a fitted parameter. But a deterministic value of $p$ (Eq. \ref{eq:TI}) provides an accurate description of electronic recoil data only for very small energies $E\lesssim2$~keV, where recombination and recombination fluctuations are small \cite{CH3T}. At higher energies, the previously mentioned excess variance manifests itself. A simple way to modify this general picture to account for the excess variance is to let $p$ itself vary, so that in Eq. \ref{eq:TI}, $p \rightarrow \langle p \rangle$. One way it can be modeled is by a Gaussian distribution with fixed width $\sigma_p \approx 0.06$ \cite{Dobi}. In terms of the notation of approach (a), $\sigma_p = \sigma_r / n_i$. 

The total variance due to the recombination process as described above is
\begin{equation}\label{eq:TV}
\begin{aligned}
  \sigma_T^2 &= \sigma_b^2 + \sigma_r^2 \\
  &= (1-p) n_i p + (\sigma_p n_i)^2, 
\end{aligned}
\end{equation}
in which $\sigma_b^2$ is the binomial variance. Eq. \ref{eq:TV} immediately shows how approach (a) is the large-$n_i$ case of the General Picture just described: for nearly all measurable event energies, $\sigma_r \gg \sigma_b$.

For a standard Poisson distribution, the expected variance from the numerical approximation in approach (b) would be $\sigma^2_{\textrm{Poiss}}=n_ip$ prior to any Gaussian modification. To satisfy fluctuations with $\sigma^2_r \propto n^2_i$, the method outlined in \cite{NEST2014} defines a factor, $\mathcal{F}_r=\sigma^2_r/\sigma^2_{\textrm{Poiss}}$. Specifically, for agreement with measurements using approach (a) in \cite{Dobi},
\begin{equation}
\mathcal{F}_r=\frac{(0.06)^2}{p}n_i.
\end{equation}
This $\mathcal{F}_r$ factor appears in the variance of the Gaussian distribution that is used to broaden the Poisson distribution. The size of the fluctuations relative to a typical Poisson distribution is parameterized with the constant, $\omega$, where
\begin{equation}
\begin{aligned}
  \sigma^2_r &=\mathcal{F}_r n_i p \\
  &=\left(\omega n_i - 1\right) n_i p \\
  &\approx \omega p n^2_i~~~\left(n_i \textrm{ large}\right).
\end{aligned}
\end{equation}
Averaging the electronic recoil escape probability for over all measurable energies ($p \approx 0.5$), the expectation with approach (b) is $\omega \approx$ 0.007 in order to match $\sigma_p \approx 0.06$~\cite{Dobi}.

It is surprising that the distribution of $p$ maintains a fixed width, independent of $n_i$ (or, in approach (b), nearly fixed width due to the $p$-dependence introduced by the Poisson approximation). A possible physical interpretation could be the initial energy distribution of ionization electrons. This would map directly into their recombination probability, in the limit of isolated electron-ion pairs. Further investigation into this hypothesis is beyond the scope of the present work.

\subsection{\label{hiErERfluct}Analysis of electronic recoils with $E\gtrsim10$~keV}
\begin{figure}
	\includegraphics[scale=0.46]{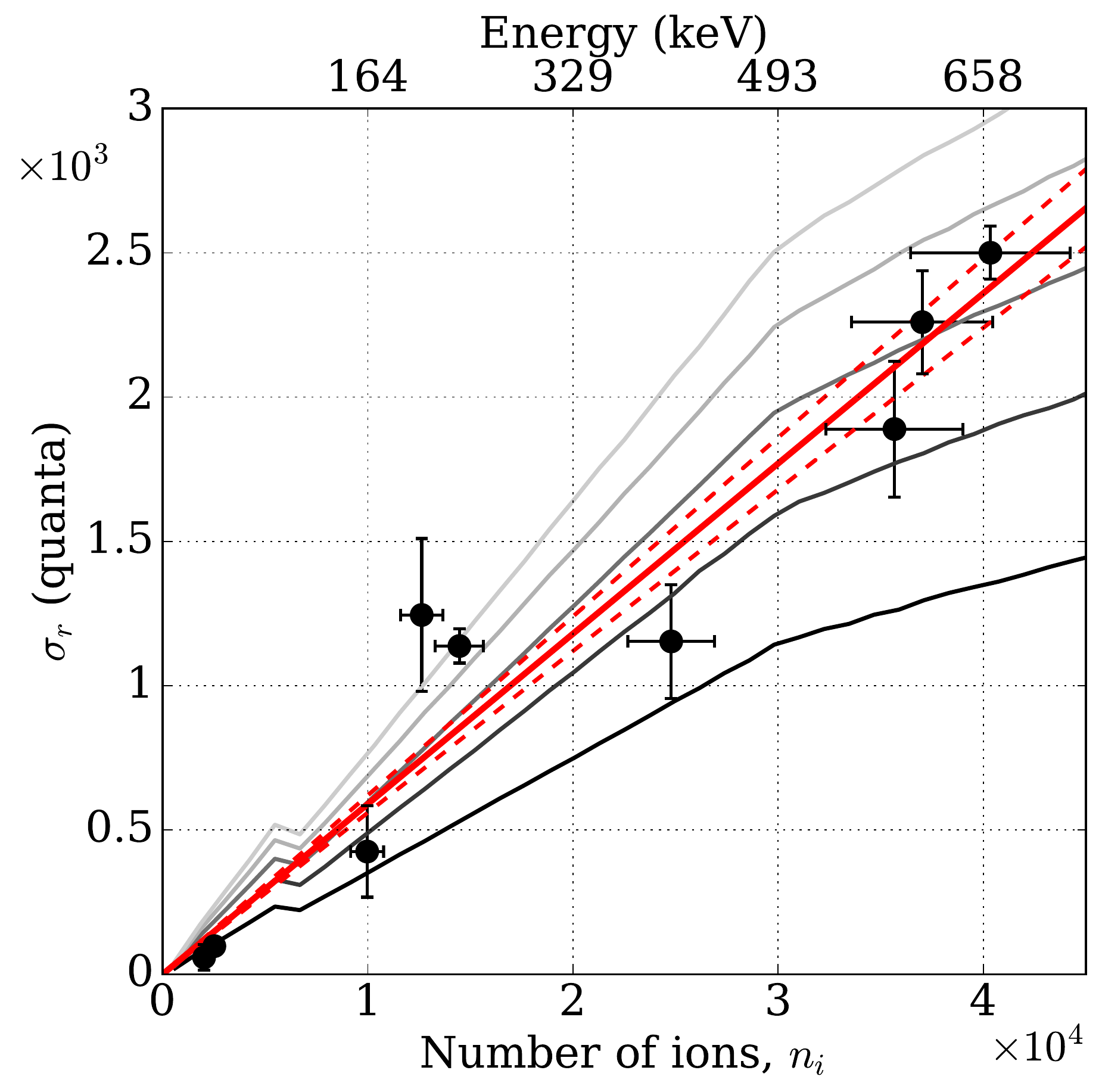}
    \caption{\label{sigmaRni} The measured Gaussian recombination fluctuations (as described in approach (a)) scale linearly with the number of ions. The solid (dashed) red lines are the linear best fit ($\pm 1\sigma$) calculated to be $\sigma_r=\left(0.059\pm0.003\right)n_i$. The other solid lines show the NEST model (approach (b)) for five choices of $\omega$, where the line color changes with $\omega$ = 0.0025 (black), 0.005, 0.0075, 0.01, 0.0125 (lightest gray).}
\end{figure}

In this section, the purely Gaussian approach (a) is pursued. The measured widths of energy, light, and charge peaks contain information on both detector resolution and physical fluctuations in the amount of recombination. Finite detector resolution broadens the S1 and S2 peaks independently. Using Figure \ref{s1s2plot} as an example, recombination fluctuations slide events along the diagonal line of constant energy (the major axis of the ellipse) exchanging quanta of light for those of charge, or vice versa. We directly measure the detector resolution for light ($\sigma_{S1}$), charge ($\sigma_{S2}$), and energy ($\sigma_{E}$), and calculate the recombination fluctuations ($\sigma_r$) following the method in \cite{Dobi}:
\begin{equation}
\sigma_r^2 = \frac{1}{2}\left(\frac{\sigma_{S1}^2}{g_1^2} + \frac{\sigma_{S2}^2}{g_2^2} - \frac{\sigma_{E}^2}{W^2}\right) .
\end{equation}
A detailed analysis of these processes at low energies from tritium beta calibrations of the LUX detector is included in \cite{CH3T}, which notes linear scaling of $\sigma_r$ with $n_i$ for energies 2 to 16~keV. A linear model also describes the recombination fluctuations measured out to 661.7 keV, $\sigma_r=(0.059\pm0.003)n_i$, shown in Fig.~\ref{sigmaRni}. The measured slope from these higher-energy data is consistent with the measurement from tritium.

\subsection{\label{ERfluct}Analysis of electronic recoils with $E\lesssim10$~keV}

\begin{figure}
	\includegraphics[scale=0.41]{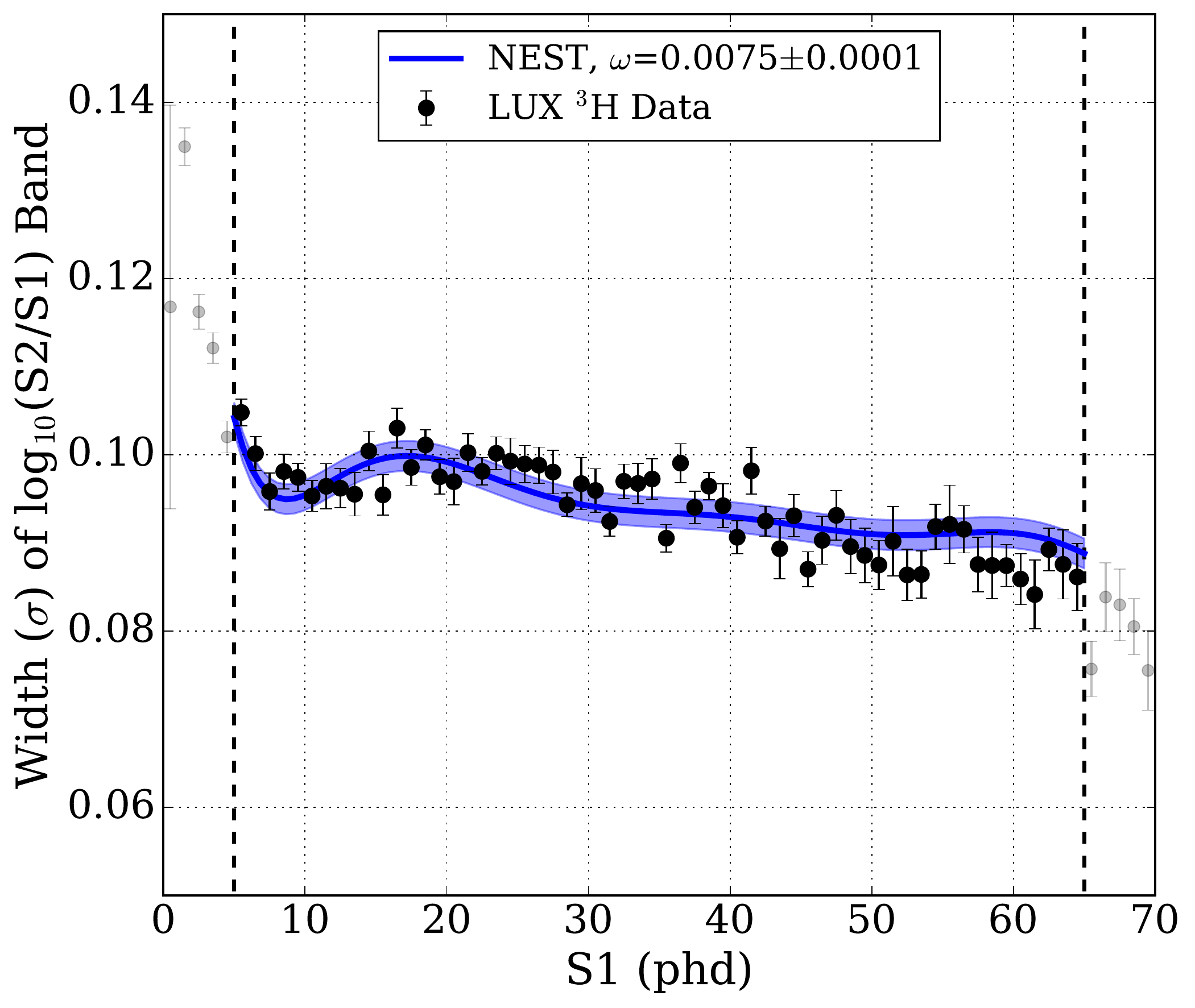}
    \caption{\label{LUXch3tBandSigma} The ER band width is plotted with the best fit NEST ER band width. Note that NEST here has been customized for these low energies as in \cite{CH3T}. The data points with full opacity are used for the fit of $\omega$. NEST's Poisson implementation of fluctuations breaks down below S1 = 5 phd, and lower $^3$H statistics above S1 = 65 phd. The solid blue line is the NEST prediction for $\omega=0.0075$, while the shaded blue region shows the variation for the range of uncertainty in $\omega$.}
\end{figure}

The low-energy ER calibration of the LUX detector was accomplished with the injection of tritiated methane. A 10 Bq injection of CH$_3$T in December 2013 produced 300,000 events in the active region with 170,000 of those occurring in the fiducial volume \cite{CH3T}. Using NEST's numerical implementation of approach (b), a $\chi^2$ comparison of simulated $^3$H electronic recoils with varying $\omega$ is made with LUX $^3$H data. As previously described, the manifestation of the variance from recombination is in the width of the $\log_{10}\left(\textrm{S2}/\textrm{S1}\right)$ signal band. The $\chi^2$ was calculated from the Gaussian width of this band in data and from the width of the same band from NEST Monte Carlo (MC) simulation, sweeping from $\omega=0.001$ to 0.011. The best fit value from this NEST MC approach is $\omega = 0.0075 \pm 0.0001$, and it is shown in Fig. 11 along with LUX $^3$H ER calibration data. This NEST result can be reconciled with the Gaussian measurement for recombination fluctuations in these same LUX data. At low energies, our numerical treatment of fluctuations cannot be strictly Gaussian because predictions of negative numbers of quanta are unphysical, but the resulting variance in observed quanta can be compared to the more straight-forward Gaussian models at higher energies. The fit parameters of interest from each approach are consistent, as $\sigma_p=0.067\pm0.005$ from the Gaussian approximation of recombination fluctuations in \cite{CH3T} is approximately equal to this NEST MC fit for $\sqrt{\omega p}$ (Eq.~12). If the average escape probability is considered for 2-10 keV, using Fig.~\ref{recombplot} from which one finds $p=0.7-0.3$, the expected value for $\sqrt{\omega p}$ is approximately $\sqrt{(0.0075\pm0.0001)\cdot(0.3-0.7)} \simeq 0.05-0.07$. This shows the consistency of multiple models' treatment of the same data, and also agreement with Sec.~\ref{hiErERfluct}, a remarkable general result for electronic recoil recombination fluctuations across orders of magnitude in deposited energy.

\begin{figure}
	\includegraphics[scale=0.41]{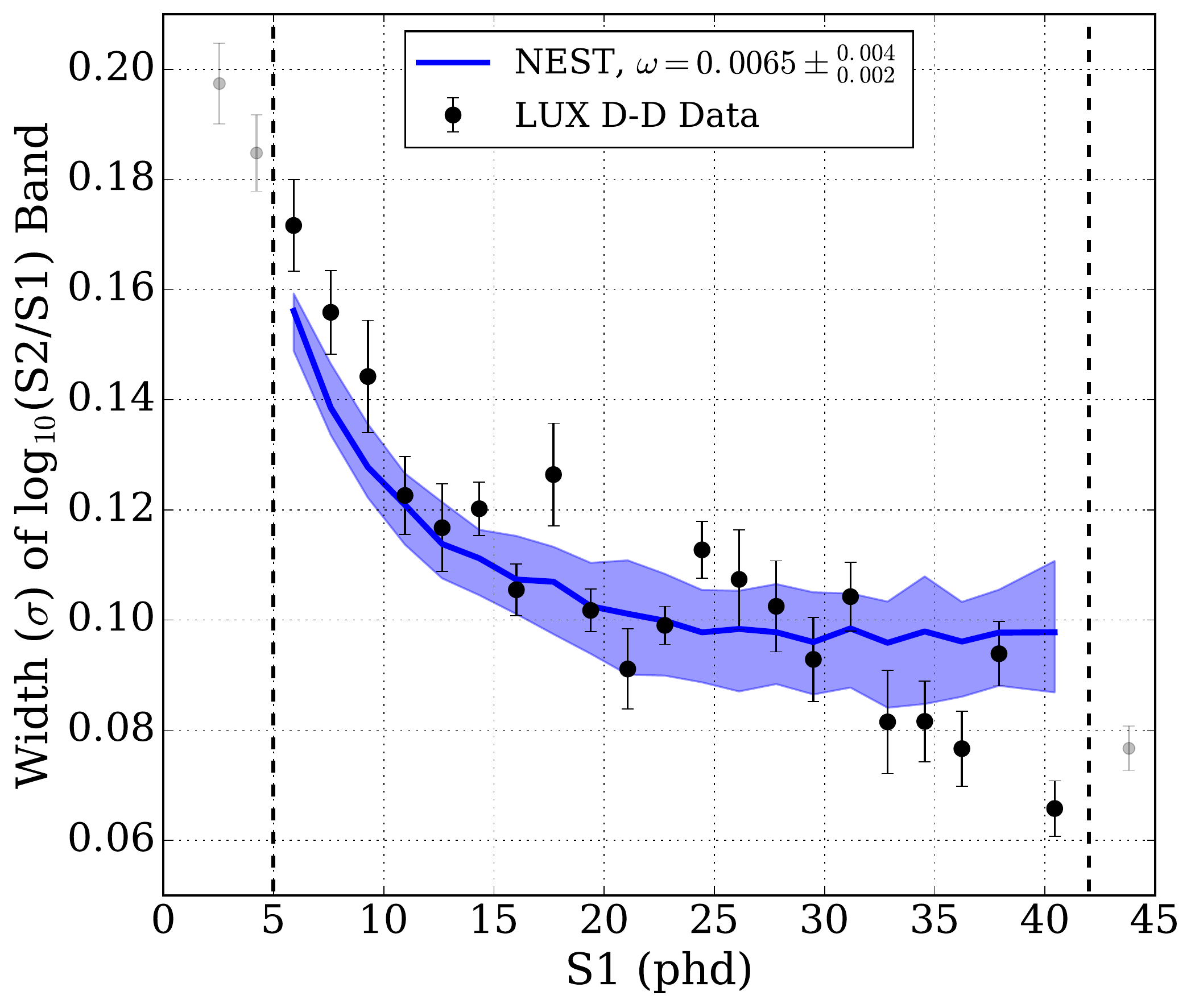}
    \caption{\label{NRsigmaplt} The NR band width is plotted with the best fit NEST NR band width. Note that NEST here is a modified version of \cite{NEST2011} as in \cite{run4}. The data points with full opacity are used for the fit of $\omega$, while the semi-transparent points are excluded due to limited statistics and large variation in the fit values from the mid-range energies of the D-D recoil spectrum with lower event rate (begins near S1 = 35-40 phd). The solid blue line is the NEST prediction for $\omega=0.0065$, while the shaded blue region shows the variation for the range of uncertainty in $\omega$.}
\end{figure}

\subsection{\label{NRfluct}Analysis of nuclear recoils}
The low-energy NR calibration of the LUX detector occurred within one month of the $^3$H ER calibration of the previous section. It was made with a collimated beam of 2.45 MeV deuterium-deuterium (D-D) neutrons made by producing the neutrons outside of the water tank and having them travel unimpeded through the tank via an air-filled tube. An appreciable fraction of the neutrons pass through the cryostat and detector materials and deposit energy in the liquid xenon with single or multiple scatters. For a detailed description of this calibration, see \cite{DDpaper}. Using the same method as in Sec. \ref{ERfluct} and exploiting the NR functionality of the NEST framework \cite{NEST2014}, the $\chi^2$ comparison yields a best fit for $\omega=0.0065\pm^{0.004}_{0.002}$. The $\log_{10}(S2/S1)$ band $\sigma$ measurements from the D-D data are plotted with the best fit from NEST in Fig.~\ref{NRsigmaplt}.

As evidenced by the large uncertainties for $\omega$ from these LUX NR data, a definitive statement cannot be made about the two descriptions of NR recombination fluctuations, one strictly binomial and the other with an $n_i^2$ term in the variance as in ER recombination. The similarity and proximity of the best-fit values for both recoil types is noteworthy. In practice, for example, the same $\omega=0.0075$ was used successfully for ER and NR models in \cite{reanalysis} and falls within the range of uncertainty. While beyond the scope this paper, this merits further study in future analyses.

\begin{table}
\begin{ruledtabular}
\begin{tabular}{ccc}
    Energy (keV) & $L_y$ (photons/keV) & $Q_y$ (electrons/keV)\\
    \hline
    5.2  & $39.2 \pm 1.9_\text{stat} \pm 1.0_\text{sys}$ & $31.0 \pm 0.6_\text{stat} \pm 2.4_\text{sys}$ \\
    33.2  & $49.5 \pm 0.4_\text{stat} \pm 1.3_\text{sys}$ & $22.9 \pm 0.3_\text{stat} \pm 1.7_\text{sys}$ \\
    41.55 & $53.4 \pm 0.0_\text{stat} \pm 1.4_\text{sys}$ & $19.4 \pm 0.0_\text{stat} \pm 1.4_\text{sys}$ \\
    163.9 & $41.9 \pm 0.3_\text{stat} \pm 1.1_\text{sys}$ & $28.3 \pm 0.9_\text{stat} \pm 2.1_\text{sys}$ \\
    208 & $43.1 \pm 0.5_\text{stat} \pm 1.1_\text{sys}$ & $29.9 \pm 0.7_\text{stat} \pm 2.3_\text{sys}$ \\
    236.1 & $43.9 \pm 0.2_\text{stat} \pm 1.1_\text{sys}$ & $29.5 \pm 0.2_\text{stat} \pm 2.3_\text{sys}$ \\
    410 & $42.4 \pm 0.3_\text{stat} \pm 1.1_\text{sys}$ & $29.7 \pm 0.5_\text{stat} \pm 2.4_\text{sys}$ \\
    583.2 & $35.5 \pm 0.3_\text{stat} \pm 0.9_\text{sys}$ & $38.0 \pm 0.2_\text{stat} \pm 3.4_\text{sys}$ \\
    609.3 & $37.4 \pm 0.1_\text{stat} \pm 1.0_\text{sys}$ & $35.2 \pm 0.2_\text{stat} \pm 3.1_\text{sys}$ \\
    661.7 & $35.1 \pm 0.1_\text{stat} \pm 0.9_\text{sys}$ & $37.7 \pm 0.1_\text{stat} \pm 3.5_\text{sys}$ \\
\end{tabular}
\end{ruledtabular}
\caption{\label{tab:lyqy} The numerical values for the plotted yields from Figures \ref{lyplot} and \ref{qyplot}.  The primary systematic uncertainties are propagated from $g_1$ and $g_2$. There is an additional energy-dependent contribution above 200 keV from variation in the amount of single-electron contamination, and a 6\% contribution from S2 DAQ saturation of measurements at 583.2, 609.3, and 661.7 keV. Statistical uncertainties are completely subdominant to systematics for measurements above 33.2 keV.}
\end{table}
\begin{table}[t]
\begin{ruledtabular}
\begin{tabular}{ccc}
    Energy (keV) & Source & Resolution $\left(\sigma / \mu\right)$\\
    \hline
    5.2 & $^{127}$Xe & $0.124 \pm 0.004_\text{stat} \pm 0.010_\text{sys}$ \\
    33.2 & $^{127}$Xe & $0.052 \pm 0.001_\text{stat} \pm 0.004_\text{sys}$ \\
    41.55 & $^{83\textrm{m}}$Kr & $0.053 \pm 0.004_\text{sys}$ \\
    163.9 & $^{131\textrm{m}}$Xe & $0.028 \pm 0.002_\text{sys}$ \\
    208 & $^{127}$Xe & $0.024 \pm 0.002_\text{sys}$ \\
    236.1 & $^{127}$Xe, $^{129\textrm{m}}$Xe & $0.026 \pm 0.002_\text{sys}$ \\
    410 & $^{127}$Xe & $0.022 \pm 0.002_\text{sys}$ \\
    583.2 & $^{208}$Tl & $0.026 \pm 0.003_\text{sys}$ \\
    609.3 & $^{214}$Bi & $0.030 \pm 0.003_\text{sys}$ \\
    661.7 & $^{137}$Cs & $0.028 \pm 0.003_\text{sys}$ \\
\end{tabular}
\end{ruledtabular}
\caption{\label{tab:epeaks} The numerical values for energy resolution as plotted in Figure \ref{fig:eresplot}. The primary systematic uncertainties are propagated from $g_1$ and $g_2$ with an energy-dependent contribution above 200 keV from variation in the amount of single-electron contamination and a 6\% contribution from S2 DAQ saturation of measurements at 583.2, 609.3, and 661.7 keV. Statistical uncertainties are completely subdominant to systematics for measurements above 33.2 keV.}
\end{table}

\section{\label{Disc}Summary}
With data from the LUX detector we have measured light and charge yields and calculated the mean recombination at the energies of many common ER background and calibration sources for LXe detectors. The light and charge yields measured with LUX are almost completely consistent within uncertainties with current NEST models of electronic recoils in LXe for energies 5 to 661.7~keV. The LUX data show a lower light yield and higher charge yield at 33.2~keV, a challenging energy to model where NEST transitions from the Thomas-Imel model to the Doke model for recombination. Composite yields from the multiple-step decays of activated xenon are consistent with the predicted quanta from multiple smaller energy deposits and distinct from the yields of a single deposition of the total energy. Measurements of the LUX energy resolution are competitive with previous measurements by smaller LXe TPCs at low energies. The degraded energy resolution at high energies is caused by known effects in the S2 channel.\\

LUX measurements of recombination fluctuations reinforce previous observations of larger-than-binomial variance. Measurements made by following two approaches that originate from the same general description of recombination help clarify the agreement between these measurements, prior measurements, and the present numerical implementation of this physics. While the general description described in Eq.~10 is likely the ``most correct," the Poisson and Gaussian approaches are necessary due to the computational expense of binomial processes for any practical use in NEST and other simulation packages. Dedicated tests of possible physical interpretations of the additional variance (e.g. electron-ion track structure) and the differences stemming from recoil type should be pursued.

\section{Acknowledgments}
This work was partially supported by the U.S. Department of Energy (DOE) under award numbers DE-AC02-05CH11231, DE-AC05-06OR23100, DE-AC52-07NA27344, DE-FG01-91ER40618, DE-FG02-08ER41549, DE-FG02-11ER41738, DE-FG02-91ER40674, DE-FG02-91ER40688, DE-FG02-95ER40917, DE-NA0000979, DE-SC0006605, DE-SC0010010, and DE-SC0015535; the U.S. National Science Foundation under award numbers PHY-0750671, PHY-0801536, PHY-1003660, PHY-1004661, PHY-1102470, PHY-1312561, PHY-1347449, PHY-1505868, and PHY-1636738; the Research Corporation grant RA0350; the Center for Ultra-low Background Experiments in the Dakotas (CUBED); and the South Dakota School of Mines and Technology (SDSMT). LIP-Coimbra acknowledges funding from Funda\c{c}\~{a}o para a Ci\^{e}ncia e a Tecnologia (FCT) through the project-grant PTDC/FIS-NUC/1525/2014. Imperial College and Brown University thank the UK Royal Society for travel funds under the International Exchange Scheme (IE120804). The UK groups acknowledge institutional support from Imperial College London, University College London and Edinburgh University, and from the Science \& Technology Facilities Council for PhD studentships ST/K502042/1 (AB), ST/K502406/1 (SS) and ST/M503538/1 (KY). The University of Edinburgh is a charitable body, registered in Scotland, with registration number SC005336.

This research was conducted using computational resources and services at the Center for Computation and Visualization, Brown University, and also the Yale Science Research Software Core. The $^{83}$Rb used in this research to produce $^{83\mathrm{m}}$Kr was supplied by the United States Department of Energy Office of Science by the Isotope Program in the Office of Nuclear Physics.

We gratefully acknowledge the logistical and technical support and the access to laboratory infrastructure provided to us by SURF and its personnel at Lead, South Dakota. SURF was developed by the South Dakota Science and Technology Authority, with an important philanthropic donation from T. Denny Sanford, and is operated by Lawrence Berkeley National Laboratory for the Department of Energy, Office of High Energy Physics.

\bibliography{main}

\begin{thebibliography}{33}%
\makeatletter
\providecommand \@ifxundefined [1]{%
 \@ifx{#1\undefined}
}%
\providecommand \@ifnum [1]{%
 \ifnum #1\expandafter \@firstoftwo
 \else \expandafter \@secondoftwo
 \fi
}%
\providecommand \@ifx [1]{%
 \ifx #1\expandafter \@firstoftwo
 \else \expandafter \@secondoftwo
 \fi
}%
\providecommand \natexlab [1]{#1}%
\providecommand \enquote  [1]{``#1''}%
\providecommand \bibnamefont  [1]{#1}%
\providecommand \bibfnamefont [1]{#1}%
\providecommand \citenamefont [1]{#1}%
\providecommand \href@noop [0]{\@secondoftwo}%
\providecommand \href [0]{\begingroup \@sanitize@url \@href}%
\providecommand \@href[1]{\@@startlink{#1}\@@href}%
\providecommand \@@href[1]{\endgroup#1\@@endlink}%
\providecommand \@sanitize@url [0]{\catcode `\\12\catcode `\$12\catcode
  `\&12\catcode `\#12\catcode `\^12\catcode `\_12\catcode `\%12\relax}%
\providecommand \@@startlink[1]{}%
\providecommand \@@endlink[0]{}%
\providecommand \url  [0]{\begingroup\@sanitize@url \@url }%
\providecommand \@url [1]{\endgroup\@href {#1}{\urlprefix }}%
\providecommand \urlprefix  [0]{URL }%
\providecommand \Eprint [0]{\href }%
\providecommand \doibase [0]{http://dx.doi.org/}%
\providecommand \selectlanguage [0]{\@gobble}%
\providecommand \bibinfo  [0]{\@secondoftwo}%
\providecommand \bibfield  [0]{\@secondoftwo}%
\providecommand \translation [1]{[#1]}%
\providecommand \BibitemOpen [0]{}%
\providecommand \bibitemStop [0]{}%
\providecommand \bibitemNoStop [0]{.\EOS\space}%
\providecommand \EOS [0]{\spacefactor3000\relax}%
\providecommand \BibitemShut  [1]{\csname bibitem#1\endcsname}%
\let\auto@bib@innerbib\@empty
\bibitem [{\citenamefont {Akerib}\ \emph {et~al.}(2013)\citenamefont {Akerib}
  \emph {et~al.}}]{NIM}%
  \BibitemOpen
  \bibfield  {author} {\bibinfo {author} {\bibfnamefont {D.~S.}\ \bibnamefont
  {Akerib}} \emph {et~al.} (\bibinfo {collaboration} {LUX Collaboration}),\
  }\href {\doibase 10.1016/j.nima.2012.11.135} {\bibfield  {journal} {\bibinfo
  {journal} {Nucl. Instrum. Methods}\ }\textbf {\bibinfo {volume} {A704}},\
  \bibinfo {pages} {111} (\bibinfo {year} {2013})}\BibitemShut {NoStop}%
\bibitem [{\citenamefont {Akerib}\ \emph {et~al.}(2014)\citenamefont {Akerib}
  \emph {et~al.}}]{PRL}%
  \BibitemOpen
  \bibfield  {author} {\bibinfo {author} {\bibfnamefont {D.~S.}\ \bibnamefont
  {Akerib}} \emph {et~al.} (\bibinfo {collaboration} {LUX Collaboration}),\
  }\href {\doibase 10.1103/PhysRevLett.112.091303} {\bibfield  {journal}
  {\bibinfo  {journal} {Phys. Rev. Lett.}\ }\textbf {\bibinfo {volume} {112}},\
  \bibinfo {pages} {091303} (\bibinfo {year} {2014})}\BibitemShut {NoStop}%
\bibitem [{\citenamefont {Akerib}\ \emph
  {et~al.}(2016{\natexlab{a}})\citenamefont {Akerib} \emph
  {et~al.}}]{reanalysis}%
  \BibitemOpen
  \bibfield  {author} {\bibinfo {author} {\bibfnamefont {D.~S.}\ \bibnamefont
  {Akerib}} \emph {et~al.} (\bibinfo {collaboration} {LUX Collaboration}),\
  }\href {\doibase 10.1103/PhysRevLett.116.161301} {\bibfield  {journal}
  {\bibinfo  {journal} {Phys. Rev. Lett.}\ }\textbf {\bibinfo {volume} {116}},\
  \bibinfo {pages} {161301} (\bibinfo {year} {2016}{\natexlab{a}})}\BibitemShut
  {NoStop}%
\bibitem [{\citenamefont {Akerib}\ \emph
  {et~al.}(2016{\natexlab{b}})\citenamefont {Akerib} \emph {et~al.}}]{run4}%
  \BibitemOpen
  \bibfield  {author} {\bibinfo {author} {\bibfnamefont {D.~S.}\ \bibnamefont
  {Akerib}} \emph {et~al.} (\bibinfo {collaboration} {LUX Collaboration}),\
  }\href@noop {} {\  (\bibinfo {year} {2016}{\natexlab{b}})},\ \Eprint
  {http://arxiv.org/abs/1608.07648} {arXiv:1608.07648 [astro-ph.CO]}
  \BibitemShut {NoStop}%
\bibitem [{\citenamefont {Akerib}\ \emph
  {et~al.}(2016{\natexlab{c}})\citenamefont {Akerib} \emph {et~al.}}]{spinDep}%
  \BibitemOpen
  \bibfield  {author} {\bibinfo {author} {\bibfnamefont {D.~S.}\ \bibnamefont
  {Akerib}} \emph {et~al.} (\bibinfo {collaboration} {LUX Collaboration}),\
  }\href {\doibase 10.1103/PhysRevLett.116.161302} {\bibfield  {journal}
  {\bibinfo  {journal} {Phys. Rev. Lett.}\ }\textbf {\bibinfo {volume} {116}},\
  \bibinfo {pages} {161302} (\bibinfo {year} {2016}{\natexlab{c}})}\BibitemShut
  {NoStop}%
\bibitem [{\citenamefont {Phelps}(2014)}]{php}%
  \BibitemOpen
  \bibfield  {author} {\bibinfo {author} {\bibfnamefont {P.}~\bibnamefont
  {Phelps}},\ }\href@noop {} {Ph.D. thesis},\ \bibinfo  {school} {Case Western
  Reserve University} (\bibinfo {year} {2014})\BibitemShut {NoStop}%
\bibitem [{\citenamefont {Akerib}\ \emph
  {et~al.}(2016{\natexlab{d}})\citenamefont {Akerib} \emph {et~al.}}]{mercury}%
  \BibitemOpen
  \bibfield  {author} {\bibinfo {author} {\bibfnamefont {D.~S.}\ \bibnamefont
  {Akerib}} \emph {et~al.} (\bibinfo {collaboration} {LUX Collaboration}),\
  }\href@noop {} {\bibfield  {journal} {\bibinfo  {journal} {Position
  reconstruction paper; in preparation}\ } (\bibinfo {year}
  {2016}{\natexlab{d}})}\BibitemShut {NoStop}%
\bibitem [{\citenamefont {Szydagis}\ \emph {et~al.}(2011)\citenamefont
  {Szydagis} \emph {et~al.}}]{NEST2011}%
  \BibitemOpen
  \bibfield  {author} {\bibinfo {author} {\bibfnamefont {M.}~\bibnamefont
  {Szydagis}} \emph {et~al.},\ }\href@noop {} {\bibfield  {journal} {\bibinfo
  {journal} {JINST}\ }\textbf {\bibinfo {volume} {6}},\ \bibinfo {pages}
  {P10002} (\bibinfo {year} {2011})}\BibitemShut {NoStop}%
\bibitem [{\citenamefont {Akerib}\ \emph
  {et~al.}(2016{\natexlab{e}})\citenamefont {Akerib} \emph {et~al.}}]{CH3T}%
  \BibitemOpen
  \bibfield  {author} {\bibinfo {author} {\bibfnamefont {D.~S.}\ \bibnamefont
  {Akerib}} \emph {et~al.} (\bibinfo {collaboration} {LUX Collaboration}),\
  }\href@noop {} {\bibfield  {journal} {\bibinfo  {journal} {Phys. Rev. D}\
  }\textbf {\bibinfo {volume} {93}},\ \bibinfo {pages} {072009} (\bibinfo
  {year} {2016}{\natexlab{e}})}\BibitemShut {NoStop}%
\bibitem [{\citenamefont {Thomas}\ and\ \citenamefont {Imel}(1987)}]{TIB}%
  \BibitemOpen
  \bibfield  {author} {\bibinfo {author} {\bibfnamefont {J.}~\bibnamefont
  {Thomas}}\ and\ \bibinfo {author} {\bibfnamefont {D.}~\bibnamefont {Imel}},\
  }\href {\doibase 10.1103/PhysRevA.36.614} {\bibfield  {journal} {\bibinfo
  {journal} {Phys. Rev. A}\ }\textbf {\bibinfo {volume} {36}},\ \bibinfo
  {pages} {614} (\bibinfo {year} {1987})}\BibitemShut {NoStop}%
\bibitem [{\citenamefont {Doke}\ \emph {et~al.}(1988)\citenamefont {Doke},
  \citenamefont {Crawford}, \citenamefont {Hitachi}, \citenamefont {Kikuchi},
  \citenamefont {Lindstrom}, \citenamefont {Masuda}, \citenamefont
  {Shibamura},\ and\ \citenamefont {Takahashi}}]{DokeRecomb}%
  \BibitemOpen
  \bibfield  {author} {\bibinfo {author} {\bibfnamefont {T.}~\bibnamefont
  {Doke}}, \bibinfo {author} {\bibfnamefont {H.}~\bibnamefont {Crawford}},
  \bibinfo {author} {\bibfnamefont {A.}~\bibnamefont {Hitachi}}, \bibinfo
  {author} {\bibfnamefont {J.}~\bibnamefont {Kikuchi}}, \bibinfo {author}
  {\bibfnamefont {P.}~\bibnamefont {Lindstrom}}, \bibinfo {author}
  {\bibfnamefont {K.}~\bibnamefont {Masuda}}, \bibinfo {author} {\bibfnamefont
  {E.}~\bibnamefont {Shibamura}}, \ and\ \bibinfo {author} {\bibfnamefont
  {T.}~\bibnamefont {Takahashi}},\ }\href {\doibase
  10.1016/0168-9002(88)90892-3} {\bibfield  {journal} {\bibinfo  {journal}
  {Nucl. Instrum. Methods}\ }\textbf {\bibinfo {volume} {A269}},\ \bibinfo
  {pages} {291} (\bibinfo {year} {1988})}\BibitemShut {NoStop}%
\bibitem [{\citenamefont {Platzman}(1961)}]{Platz}%
  \BibitemOpen
  \bibfield  {author} {\bibinfo {author} {\bibfnamefont {R.}~\bibnamefont
  {Platzman}},\ }\href {\doibase
  http://dx.doi.org/10.1016/0020-708X(61)90108-9} {\bibfield  {journal}
  {\bibinfo  {journal} {Int. J. Appl. Radiat. Is.}\ }\textbf {\bibinfo {volume}
  {10}},\ \bibinfo {pages} {116 } (\bibinfo {year} {1961})}\BibitemShut
  {NoStop}%
\bibitem [{\citenamefont {Dahl}(2009)}]{Dahl}%
  \BibitemOpen
  \bibfield  {author} {\bibinfo {author} {\bibfnamefont {C.}~\bibnamefont
  {Dahl}},\ }\href@noop {} {Ph.D. thesis},\ \bibinfo  {school} {Princeton
  University} (\bibinfo {year} {2009})\BibitemShut {NoStop}%
\bibitem [{\citenamefont {Akerib}\ \emph
  {et~al.}(2016{\natexlab{f}})\citenamefont {Akerib} \emph {et~al.}}]{DDpaper}%
  \BibitemOpen
  \bibfield  {author} {\bibinfo {author} {\bibfnamefont {D.~S.}\ \bibnamefont
  {Akerib}} \emph {et~al.} (\bibinfo {collaboration} {LUX Collaboration}),\
  }\href@noop {} {\  (\bibinfo {year} {2016}{\natexlab{f}})},\ \Eprint
  {http://arxiv.org/abs/1608.05381} {arXiv:1608.05381 [physics.ins-det]}
  \BibitemShut {NoStop}%
\bibitem [{\citenamefont {Doke}\ \emph {et~al.}(2002)\citenamefont {Doke},
  \citenamefont {Hitachi}, \citenamefont {Kikuchi}, \citenamefont {Masuda},
  \citenamefont {Okada},\ and\ \citenamefont {Shibamura}}]{exion1}%
  \BibitemOpen
  \bibfield  {author} {\bibinfo {author} {\bibfnamefont {T.}~\bibnamefont
  {Doke}}, \bibinfo {author} {\bibfnamefont {A.}~\bibnamefont {Hitachi}},
  \bibinfo {author} {\bibfnamefont {J.}~\bibnamefont {Kikuchi}}, \bibinfo
  {author} {\bibfnamefont {K.}~\bibnamefont {Masuda}}, \bibinfo {author}
  {\bibfnamefont {H.}~\bibnamefont {Okada}}, \ and\ \bibinfo {author}
  {\bibfnamefont {E.}~\bibnamefont {Shibamura}},\ }\href
  {http://stacks.iop.org/1347-4065/41/i=3R/a=1538} {\bibfield  {journal}
  {\bibinfo  {journal} {Jpn. J. Appl. Phys.}\ }\textbf {\bibinfo {volume}
  {41}},\ \bibinfo {pages} {1538} (\bibinfo {year} {2002})}\BibitemShut
  {NoStop}%
\bibitem [{\citenamefont {Aprile}\ \emph {et~al.}(2007)\citenamefont {Aprile},
  \citenamefont {Giboni}, \citenamefont {Majewski}, \citenamefont {Ni},\ and\
  \citenamefont {Yamashita}}]{exion2}%
  \BibitemOpen
  \bibfield  {author} {\bibinfo {author} {\bibfnamefont {E.}~\bibnamefont
  {Aprile}}, \bibinfo {author} {\bibfnamefont {K.}~\bibnamefont {Giboni}},
  \bibinfo {author} {\bibfnamefont {P.}~\bibnamefont {Majewski}}, \bibinfo
  {author} {\bibfnamefont {K.}~\bibnamefont {Ni}}, \ and\ \bibinfo {author}
  {\bibfnamefont {M.}~\bibnamefont {Yamashita}},\ }\href {\doibase
  10.1103/PhysRevB.76.014115} {\bibfield  {journal} {\bibinfo  {journal} {Phys.
  Rev. B}\ }\textbf {\bibinfo {volume} {76}},\ \bibinfo {pages} {014115}
  (\bibinfo {year} {2007})}\BibitemShut {NoStop}%
\bibitem [{\citenamefont {Mock}\ \emph {et~al.}(2014)\citenamefont {Mock},
  \citenamefont {Barry}, \citenamefont {Kazkaz}, \citenamefont {Stolp},
  \citenamefont {Szydagis}, \citenamefont {Tripathi}, \citenamefont {Uvarov},
  \citenamefont {Walsh},\ and\ \citenamefont {Woods}}]{Mock}%
  \BibitemOpen
  \bibfield  {author} {\bibinfo {author} {\bibfnamefont {J.}~\bibnamefont
  {Mock}}, \bibinfo {author} {\bibfnamefont {N.}~\bibnamefont {Barry}},
  \bibinfo {author} {\bibfnamefont {K.}~\bibnamefont {Kazkaz}}, \bibinfo
  {author} {\bibfnamefont {D.}~\bibnamefont {Stolp}}, \bibinfo {author}
  {\bibfnamefont {M.}~\bibnamefont {Szydagis}}, \bibinfo {author}
  {\bibfnamefont {M.}~\bibnamefont {Tripathi}}, \bibinfo {author}
  {\bibfnamefont {S.}~\bibnamefont {Uvarov}}, \bibinfo {author} {\bibfnamefont
  {N.}~\bibnamefont {Walsh}}, \ and\ \bibinfo {author} {\bibfnamefont
  {M.}~\bibnamefont {Woods}},\ }\href@noop {} {\bibfield  {journal} {\bibinfo
  {journal} {JINST}\ }\textbf {\bibinfo {volume} {9}},\ \bibinfo {pages}
  {T04002} (\bibinfo {year} {2014})}\BibitemShut {NoStop}%
\bibitem [{\citenamefont {Kubota}\ \emph {et~al.}(1978)\citenamefont {Kubota},
  \citenamefont {Hishida},\ and\ \citenamefont {Raun}}]{XePhys1}%
  \BibitemOpen
  \bibfield  {author} {\bibinfo {author} {\bibfnamefont {S.}~\bibnamefont
  {Kubota}}, \bibinfo {author} {\bibfnamefont {M.}~\bibnamefont {Hishida}}, \
  and\ \bibinfo {author} {\bibfnamefont {J.}~\bibnamefont {Raun}},\ }\href@noop
  {} {\bibfield  {journal} {\bibinfo  {journal} {J. Phys. C}\ }\textbf
  {\bibinfo {volume} {11}},\ \bibinfo {pages} {2645} (\bibinfo {year}
  {1978})}\BibitemShut {NoStop}%
\bibitem [{\citenamefont {Hitachi}\ \emph {et~al.}(1983)\citenamefont
  {Hitachi}, \citenamefont {Takahashi}, \citenamefont {Funayama}, \citenamefont
  {Masuda}, \citenamefont {Kikuchi},\ and\ \citenamefont {Doke}}]{XePhys2}%
  \BibitemOpen
  \bibfield  {author} {\bibinfo {author} {\bibfnamefont {A.}~\bibnamefont
  {Hitachi}}, \bibinfo {author} {\bibfnamefont {T.}~\bibnamefont {Takahashi}},
  \bibinfo {author} {\bibfnamefont {N.}~\bibnamefont {Funayama}}, \bibinfo
  {author} {\bibfnamefont {K.}~\bibnamefont {Masuda}}, \bibinfo {author}
  {\bibfnamefont {J.}~\bibnamefont {Kikuchi}}, \ and\ \bibinfo {author}
  {\bibfnamefont {T.}~\bibnamefont {Doke}},\ }\href {\doibase
  10.1103/PhysRevB.27.5279} {\bibfield  {journal} {\bibinfo  {journal} {Phys.
  Rev. B}\ }\textbf {\bibinfo {volume} {27}},\ \bibinfo {pages} {5279}
  (\bibinfo {year} {1983})}\BibitemShut {NoStop}%
\bibitem [{\citenamefont {Conti}\ \emph {et~al.}(2003)\citenamefont {Conti}
  \emph {et~al.}}]{Conti}%
  \BibitemOpen
  \bibfield  {author} {\bibinfo {author} {\bibfnamefont {E.}~\bibnamefont
  {Conti}} \emph {et~al.},\ }\href {\doibase 10.1103/PhysRevB.68.054201}
  {\bibfield  {journal} {\bibinfo  {journal} {Phys. Rev. B}\ }\textbf {\bibinfo
  {volume} {68}},\ \bibinfo {pages} {054201} (\bibinfo {year}
  {2003})}\BibitemShut {NoStop}%
\bibitem [{\citenamefont {Lin}\ \emph {et~al.}(2015)\citenamefont {Lin},
  \citenamefont {Fei}, \citenamefont {Gao}, \citenamefont {Hu}, \citenamefont
  {Wei}, \citenamefont {Xiao}, \citenamefont {Wang},\ and\ \citenamefont
  {Ni}}]{Kaixuan}%
  \BibitemOpen
  \bibfield  {author} {\bibinfo {author} {\bibfnamefont {Q.}~\bibnamefont
  {Lin}}, \bibinfo {author} {\bibfnamefont {J.}~\bibnamefont {Fei}}, \bibinfo
  {author} {\bibfnamefont {F.}~\bibnamefont {Gao}}, \bibinfo {author}
  {\bibfnamefont {J.}~\bibnamefont {Hu}}, \bibinfo {author} {\bibfnamefont
  {Y.}~\bibnamefont {Wei}}, \bibinfo {author} {\bibfnamefont {X.}~\bibnamefont
  {Xiao}}, \bibinfo {author} {\bibfnamefont {H.}~\bibnamefont {Wang}}, \ and\
  \bibinfo {author} {\bibfnamefont {K.}~\bibnamefont {Ni}},\ }\href {\doibase
  10.1103/PhysRevD.92.032005} {\bibfield  {journal} {\bibinfo  {journal} {Phys.
  Rev.}\ }\textbf {\bibinfo {volume} {D92}},\ \bibinfo {pages} {032005}
  (\bibinfo {year} {2015})}\BibitemShut {NoStop}%
\bibitem [{\citenamefont {Kastens}\ \emph {et~al.}(2010)\citenamefont
  {Kastens}, \citenamefont {Bedikian}, \citenamefont {Cahn}, \citenamefont
  {Manzur},\ and\ \citenamefont {McKinsey}}]{Kastens:JINST}%
  \BibitemOpen
  \bibfield  {author} {\bibinfo {author} {\bibfnamefont {L.~W.}\ \bibnamefont
  {Kastens}}, \bibinfo {author} {\bibfnamefont {S.}~\bibnamefont {Bedikian}},
  \bibinfo {author} {\bibfnamefont {S.~B.}\ \bibnamefont {Cahn}}, \bibinfo
  {author} {\bibfnamefont {A.}~\bibnamefont {Manzur}}, \ and\ \bibinfo {author}
  {\bibfnamefont {D.~N.}\ \bibnamefont {McKinsey}},\ }\href
  {http://stacks.iop.org/1748-0221/5/i=05/a=P05006} {\bibfield  {journal}
  {\bibinfo  {journal} {JINST}\ }\textbf {\bibinfo {volume} {5}},\ \bibinfo
  {pages} {P05006} (\bibinfo {year} {2010})}\BibitemShut {NoStop}%
\bibitem [{\citenamefont {Manalaysay}\ \emph {et~al.}(2010)\citenamefont
  {Manalaysay} \emph {et~al.}}]{Manalaysay:2009yq}%
  \BibitemOpen
  \bibfield  {author} {\bibinfo {author} {\bibfnamefont {A.}~\bibnamefont
  {Manalaysay}} \emph {et~al.},\ }\href {\doibase 10.1063/1.3436636} {\bibfield
   {journal} {\bibinfo  {journal} {Rev. Sci. Instrum.}\ }\textbf {\bibinfo
  {volume} {81}},\ \bibinfo {pages} {073303} (\bibinfo {year}
  {2010})}\BibitemShut {NoStop}%
\bibitem [{\citenamefont {Akerib}\ \emph
  {et~al.}(2016{\natexlab{g}})\citenamefont {Akerib} \emph {et~al.}}]{DQ127}%
  \BibitemOpen
  \bibfield  {author} {\bibinfo {author} {\bibfnamefont {D.~S.}\ \bibnamefont
  {Akerib}} \emph {et~al.} (\bibinfo {collaboration} {LUX Collaboration}),\
  }\href@noop {} {\bibfield  {journal} {\bibinfo  {journal} {$^{127}$Xe paper;
  in preparation}\ } (\bibinfo {year} {2016}{\natexlab{g}})}\BibitemShut
  {NoStop}%
\bibitem [{\citenamefont {Akerib}\ \emph
  {et~al.}(2016{\natexlab{h}})\citenamefont {Akerib} \emph {et~al.}}]{LUX:PRD}%
  \BibitemOpen
  \bibfield  {author} {\bibinfo {author} {\bibfnamefont {D.~S.}\ \bibnamefont
  {Akerib}} \emph {et~al.} (\bibinfo {collaboration} {LUX Collaboration}),\
  }\href@noop {} {\bibfield  {journal} {\bibinfo  {journal} {LUX2013
  comprehensive paper; in preparation}\ } (\bibinfo {year}
  {2016}{\natexlab{h}})}\BibitemShut {NoStop}%
\bibitem [{\citenamefont {DeStefano}(2015)}]{PIXeYthesis}%
  \BibitemOpen
  \bibfield  {author} {\bibinfo {author} {\bibfnamefont {N.}~\bibnamefont
  {DeStefano}},\ }\href@noop {} {Ph.D. thesis},\ \bibinfo  {school} {University
  of Connecticut} (\bibinfo {year} {2015})\BibitemShut {NoStop}%
\bibitem [{\citenamefont {DeStefano}\ \emph {et~al.}(2016)\citenamefont
  {DeStefano} \emph {et~al.}}]{PIXeY}%
  \BibitemOpen
  \bibfield  {author} {\bibinfo {author} {\bibfnamefont {N.}~\bibnamefont
  {DeStefano}} \emph {et~al.} (\bibinfo {collaboration} {PIXeY
  Collaboration}),\ }\href@noop {} {\bibfield  {journal} {\bibinfo  {journal}
  {in preparation}\ } (\bibinfo {year} {2016})}\BibitemShut {NoStop}%
\bibitem [{\citenamefont {Stephenson}\ \emph {et~al.}(2015)\citenamefont
  {Stephenson} \emph {et~al.}}]{MiX}%
  \BibitemOpen
  \bibfield  {author} {\bibinfo {author} {\bibfnamefont {S.}~\bibnamefont
  {Stephenson}} \emph {et~al.},\ }\href {\doibase
  10.1088/1748-0221/10/10/P10040} {\bibfield  {journal} {\bibinfo  {journal}
  {JINST}\ }\textbf {\bibinfo {volume} {10}},\ \bibinfo {pages} {P10040}
  (\bibinfo {year} {2015})}\BibitemShut {NoStop}%
\bibitem [{\citenamefont {Akimov}\ \emph {et~al.}(2012)\citenamefont {Akimov}
  \emph {et~al.}}]{Akimov}%
  \BibitemOpen
  \bibfield  {author} {\bibinfo {author} {\bibfnamefont {D.}~\bibnamefont
  {Akimov}} \emph {et~al.} (\bibinfo {collaboration} {ZEPLIN-III
  Collaboration}),\ }\href {\doibase
  http://dx.doi.org/10.1016/j.physletb.2012.01.064} {\bibfield  {journal}
  {\bibinfo  {journal} {Phys. Lett. B}\ }\textbf {\bibinfo {volume} {709}},\
  \bibinfo {pages} {14 } (\bibinfo {year} {2012})}\BibitemShut {NoStop}%
\bibitem [{\citenamefont {Aprile}\ \emph {et~al.}(2012)\citenamefont {Aprile}
  \emph {et~al.}}]{Xenon100}%
  \BibitemOpen
  \bibfield  {author} {\bibinfo {author} {\bibfnamefont {E.}~\bibnamefont
  {Aprile}} \emph {et~al.} (\bibinfo {collaboration} {XENON100
  Collaboration}),\ }\href@noop {} {\bibfield  {journal} {\bibinfo  {journal}
  {Astropart. Phys.}\ }\textbf {\bibinfo {volume} {35}},\ \bibinfo {pages}
  {573} (\bibinfo {year} {2012})}\BibitemShut {NoStop}%
\bibitem [{\citenamefont {Santos}\ \emph {et~al.}(2011)\citenamefont {Santos}
  \emph {et~al.}}]{zeplinSantos}%
  \BibitemOpen
  \bibfield  {author} {\bibinfo {author} {\bibfnamefont {E.}~\bibnamefont
  {Santos}} \emph {et~al.} (\bibinfo {collaboration} {ZEPLIN-III
  Collaboration}),\ }\href {\doibase 10.1007/JHEP12(2011)115} {\bibfield
  {journal} {\bibinfo  {journal} {JHEP}\ }\textbf {\bibinfo {volume} {12}},\
  \bibinfo {pages} {115} (\bibinfo {year} {2011})}\BibitemShut {NoStop}%
\bibitem [{\citenamefont {Dobi}(2014)}]{Dobi}%
  \BibitemOpen
  \bibfield  {author} {\bibinfo {author} {\bibfnamefont {A.}~\bibnamefont
  {Dobi}},\ }\href@noop {} {Ph.D. thesis},\ \bibinfo  {school} {University of
  Maryland} (\bibinfo {year} {2014})\BibitemShut {NoStop}%
\bibitem [{\citenamefont {Lenardo}\ \emph {et~al.}(2015)\citenamefont
  {Lenardo}, \citenamefont {Kazkaz}, \citenamefont {Manalaysay}, \citenamefont
  {Mock}, \citenamefont {Szydagis},\ and\ \citenamefont {Tripathi}}]{NEST2014}%
  \BibitemOpen
  \bibfield  {author} {\bibinfo {author} {\bibfnamefont {B.}~\bibnamefont
  {Lenardo}}, \bibinfo {author} {\bibfnamefont {K.}~\bibnamefont {Kazkaz}},
  \bibinfo {author} {\bibfnamefont {A.}~\bibnamefont {Manalaysay}}, \bibinfo
  {author} {\bibfnamefont {J.}~\bibnamefont {Mock}}, \bibinfo {author}
  {\bibfnamefont {M.}~\bibnamefont {Szydagis}}, \ and\ \bibinfo {author}
  {\bibfnamefont {M.}~\bibnamefont {Tripathi}},\ }\href {\doibase
  10.1109/TNS.2015.2481322} {\bibfield  {journal} {\bibinfo  {journal} {IEEE
  Trans. Nucl. Sci.}\ }\textbf {\bibinfo {volume} {62}},\ \bibinfo {pages}
  {3387} (\bibinfo {year} {2015})}\BibitemShut {NoStop}%
\end{thebibliography}%

\end{document}